\documentclass[11pt,a4paper]{article}
\pdfoutput=1
\usepackage[english]{babel}
\usepackage[latin1]{inputenc}
\usepackage{amsfonts,amsbsy,bm,euscript,mathrsfs}
\usepackage{amssymb,stmaryrd,faktor,slashed}
\usepackage[x11names]{xcolor}
\usepackage[tbtags]{amsmath}

\usepackage{graphicx}
\usepackage{cancel}
\usepackage{definitions}
\usepackage{subfigure}

\usepackage[bookmarks=true,colorlinks=true,linkcolor=black,citecolor=black,urlcolor=black,bookmarksnumbered]{hyperref}
\usepackage[nosort]{cite}

\def\NF{{N_f}}
\def\NA{{N_a}}

\DeclareMathAlphabet{\mathpzc}{OT1}{pzc}{m}{it}
\DeclareMathAlphabet{\mathcalligra}{T1}{calligra}{m}{n}

\numberwithin{equation}{subsection}

\makeatletter
\renewcommand\section{\@startsection {section}{1}{\z@}
{-3.5ex \@plus -1ex \@minus -.2ex}
{2.3ex \@plus.2ex}
{\normalfont\Large\bfseries}}
\renewcommand\subsection{\@startsection{subsection}{2}{\z@}
{-3.25ex\@plus -1ex \@minus -.2ex}
{1.5ex \@plus.2ex}
{\normalfont\large\bfseries}}
\makeatother

\def\ads{{\rm AdS}_5\times {\rm S}^5}

\begin{document}

\setcounter{equation}{0}
\setcounter{footnote}{0}
\setcounter{section}{0}

\thispagestyle{empty}

\begin{flushright} \texttt{HU-EP-16/07\\ HU-MATH-16/03}\end{flushright}

\begin{center}
\vspace{1.5truecm}

{\LARGE \bf Introduction to the thermodynamic Bethe ansatz}

\vspace{1.5truecm}

{Stijn J. van Tongeren}

\vspace{1.0truecm}

{\em Institut f\"ur Mathematik und Institut f\"ur Physik, Humboldt-Universit\"at zu Berlin, \\ IRIS Geb\"aude, Zum Grossen Windkanal 6, 12489 Berlin, Germany}

\vspace{1.0truecm}

{{\tt svantongeren (at) physik.hu-berlin.de // s.j.vantongeren (at) gmail.com}}

\vspace{1.0truecm}
\end{center}

\begin{abstract}
We give a pedagogical introduction to the thermodynamic Bethe ansatz, a method that allows us to describe the thermodynamics of integrable models whose spectrum is found via the (asymptotic) Bethe ansatz. We set the stage by deriving the Fermi-Dirac distribution and associated free energy of free electrons, and then in a similar though technically more complicated fashion treat the thermodynamics of integrable models, focusing on the one dimensional Bose gas with delta function interaction as a clean pedagogical example, secondly the XXX spin chain as an elementary (lattice) model with prototypical complicating features in the form of bound states, and finally the $\mathrm{SU}(2)$ chiral Gross-Neveu model as a field theory example. Throughout this discussion we emphasize the central role of particle and hole densities, whose relations determine the model under consideration. We then discuss tricks that allow us to use the same methods to describe the exact spectra of integrable field theories on a circle, in particular the chiral Gross-Neveu model. We moreover discuss the simplification of TBA equations to Y systems, including the transition back to integral equations given sufficient analyticity data, in simple examples.
\end{abstract}

\newpage

\setcounter{equation}{0}
\setcounter{footnote}{0}
\setcounter{section}{0}

\tableofcontents

\section{Introduction}

Integrable models are an important class of physical models because they are ``solvable'' -- meaning we can often exactly compute various quantities -- while sharing important features with more complicated physical models. In other words, they make great pedagogical tools. Integrability makes it possible to diagonalize the chiral Gross-Neveu model's Hamiltonian for instance \cite{Andrei:1979sq,Belavin:1979pq}, giving exact formulas that explicitly demonstrate deep quantum field theoretical concepts such as dimensional transmutation and asymptotic freedom. As part of a series of articles introducing aspects of integrability \cite{Bombardelli:2016rwb}, in this article we describe how integrability is used to describe the exact thermodynamics of integrable models, and relatedly the spectra of integrable field theories defined on a circle, using a method known as the ``thermodynamic Bethe ansatz''.

As the name implies, the thermodynamic Bethe ansatz (TBA) revolves around applying the Bethe ansatz in a thermodynamic setting. In essence, the Bethe ansatz description of an integrable model provides us with momenta and energy distributions of particles, which in principle contains the information needed to determine the density of states in the thermodynamic limit, and the associated particle and hole distributions in thermodynamic equilibrium. This approach was pioneered in the late sixties by Yang and Yang \cite{Yang:1968rm} who applied it to the Bose gas with delta function interaction, also known as the Lieb-Liniger model \cite{Lieb:1963rt}. It was quickly adapted to lattice integrable models such as the Heisenberg spin chain \cite{Gaudin:1971gt,Takahashi:1972,Takahashi:book} and Hubbard model \cite{Takahashi:1972hub,Korepin}.\footnote{While we aim to focus on the basic structure, the TBA and related methods also play an important role in computing more complicated observables such as correlation functions at finite temperature, see e.g. \cite{Gohmann:2004ja}.} The TBA can be used to compute the free energy of integrable field theories as well, which upon doing a double Wick rotation has an alternative use in finding their exact ground state energies in finite volume \cite{Zamolodchikov:1989cf}. By a form of analytic continuation excited state energies can also be computed in the TBA approach \cite{Dorey:1996re,Bazhanov:1996aq}. These equations can be simplified and reduced to a so-called Y system \cite{Zamolodchikov:1991et}, which is a set of functional relations not limited to a particular state which can be the same for different models. Providing a sufficient amount of analyticity data then singles out a model and state.\footnote{Going a bit beyond the scope of the present article, such Y systems together with analyticity data can be ``reduced'' even further via so-called T systems to Q systems. Sometimes we can derive such functional relations by direct computations in a model, which can then be turned into integral equations possibly of TBA type. This comes back in the article by S. Negro \cite{Negro:2016yuu}.}

In the context of the AdS/CFT correspondence, the worldsheet theory of the $\ads$ string is an integrable field theory, see e.g. \cite{Arutyunov:2009ga,Beisert:2010jr} for reviews, and its exact energy spectrum can be computed by means of the thermodynamic Bethe ansatz \cite{Arutyunov:2007tc,Arutyunov:2009zu,Arutyunov:2009ur,Bombardelli:2009ns,Gromov:2009bc}, as first suggested in \cite{Ambjorn:2005wa}.\footnote{In this context the Y system was conjectured in \cite{Gromov:2009tv} and the required analyticity data clarified in \cite{Cavaglia:2010nm,Cavaglia:2011kd,Balog:2011nm}. Reducing this results in a Q system, in this context dubbed the quantum spectral curve \cite{Gromov:2013pga}.} This energy spectrum is AdS/CFT dual to the spectrum of scaling dimensions in planar $\mathcal{N}=4$ supersymmetric Yang-Mills theory (SYM). Provided we take the AdS/CFT correspondence to hold rigorously, the thermodynamic Bethe ansatz therefore allows us to find exact two point functions in an interacting, albeit planar, four dimensional quantum field theory, nonperturbatively. From a different point of view, this approach provides high precision tests of the AdS/CFT conjecture. The TBA approach has for instance been successfully matched by explicit field theory results up to five loops for the so-called Konishi operator \cite{Bajnok:2009vm,Arutyunov:2010gb,Balog:2010xa,Eden:2012fe}. The TBA can also be used to compute the generalized cusp anomalous dimension (the ``quark--anti-quark potential'') \cite{Drukker:2012de,Correa:2012hh}, and for instance extends to the duality between strings on the Lunin-Maldacena background and $\beta$ deformed SYM \cite{Gromov:2010dy,Arutyunov:2010gu} and the $\mathrm{AdS}_4 \times \mathbb{CP}^3$ string dual to three dimensional $\mathcal{N}=6$ supersymmetric Chern-Simons theory \cite{Bombardelli:2009xz,Gromov:2009at}. Though TBA-like equations have not yet made a clear appearance in the computation of three point correlation functions in SYM, we can expect they will do so in the exact solution.

Taking in the above, our motivation for studying the TBA is therefore broadly speaking twofold: with it we can describe the thermodynamics of nontrivial interacting models of for instance magnetism and strongly correlated electrons of relevance in condensed matter physics, as well as the exact spectra of integrable field theories that play an important role in for example string theory and the gauge/gravity duality. We will not aim to describe the technical details required for particular applications. Rather, we will focus on the unifying features of the TBA approach, and explain them such that it is clear where and how details of a particular model are to be inserted. We will nevertheless use concrete examples, first of all the original case of the Bose gas as a particularly clean example where the transition from Bethe ansatz to thermodynamic Bethe ansatz is a fairly rigorous derivation. We will also discuss the XXX Heisenberg magnet in the context of spin chains, and the $\mathrm{SU}(2)$ chiral Gross-Neveu model in integrable field theory. These models illustrate complicating hypotheses in the TBA approach to general integrable models: the presence of multiple interacting particle species, as well as bound state solutions.

We will begin our discussion with free electrons, a trivially integrable model, where we can link our approach to standard statistical physics. This allows us to introduce the concept of density of states, particle and hole density, and the computation of the associated free energy, and reproduce the well known Fermi-Dirac distribution. Following Yang and Yang's original paper, we then extend this framework to the delta function Bose gas. Continuing to the XXX spin chain and $\mathrm{SU}(2)$ chiral Gross-Neveu model in the same spirit, requires us to introduce the so-called string hypothesis, and ultimately results in an infinite set of TBA equations. We discuss how these TBA equations can be ``simplified'' and reduced to a so-called Y system. Next we discuss the TBA approach to exact ground state energies, and indicate how excited state TBA equations can be obtained by analytic continuation, motivated by a toy model example. Relatedly, we discuss the link between the TBA equations and so-called L\"uscher corrections, providing analyticity data for excited states. We briefly discuss universality of the Y system for excited states, how to transfer between TBA and Y system plus analyticity data, and the relation of the analyticity data to specific models and states. Two appendices contain details on integral identities and some comments on numerically solving TBA equations.

\section{The thermodynamic Bethe ansatz}

In an integrable model we usually have a set of Bethe ansatz equations that determines the momenta of particles of any state of the theory, either exactly, or approximately in a large volume limit. In what follows we will assume these to be given, for instance following the discussion in \cite{Levkovich-Maslyuk:2016kfv}. Combining these Bethe equations with the dispersion relation of the theory under consideration, we can determine its (approximate) energy spectrum. What if we are interested in the thermodynamic limit? Since we can in principle determine the possible and actual momentum distributions of particles for \emph{any} given set of finite quantum numbers (at large volume), we might be able to determine nontrivial thermodynamic quantities by summing up many contributions. The technical way to do this goes under the name of the thermodynamic Bethe ansatz, as originally developed by Yang and Yang for the one dimensional Bose gas with delta function interaction potential \cite{Yang:1968rm}. We will get to this model and the chiral Gross-Neveu model shortly, but let us begin with a trivially integrable model: free electrons. Our discussion will be similar to section 5.1 of \cite{Korepin}.

\subsection{Free Fermi gas}

\begin{figure}%
\centering
\includegraphics{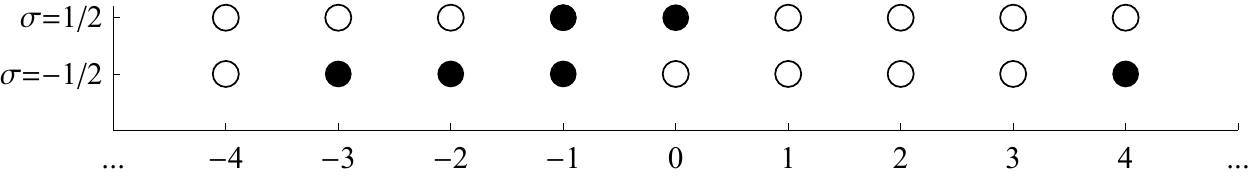}
\caption{The quantum number lattice for electrons. States of an $N$ free electron state on a circle can be labeled by a set of $N$ integers, split in sets of distinct ones for each spin. Here these integers are represented by filled dots, open dots representing available (unoccupied) quantum states, depicting a state with two spin up electrons and four spin down electrons, with momenta $-2\pi/L$ and $0$, and $-6\pi/L,-4\pi/L,-2\pi/L,$ and $8\pi/L$ respectively.}
\label{fig:freeelectronnumbers}
\end{figure}

Free electrons on a circle are an exactly solvable model. Since the particles do not interact (except for Pauli exclusion), wavefunctions are just superpositions of standing waves on the circle, each coming with a momentum quantization condition
\begin{equation}
e^{i p_j L} = 1 \implies p_j = \frac{2\pi n_j}{L} .
\end{equation}
Were we to consider fermions on a periodic lattice (with spacing one), mode numbers would of course only be meaningful modulo $L$. The Pauli exclusion principle now simply requires that each state is made up of electrons with distinct sets of quantum numbers (including spin). Note that the above equations are nothing but the simplest of Bethe equations. In fact, you might recall that in the Bethe ansatz two identical particles by construction cannot have equal momenta either, which is why we are looking at free fermions rather than free bosons. An $N$ particle state can now be classified by $N$ quantum numbers $n_j$, split in two sets $\{n^\sigma_j\}$ of distinct numbers, where $\sigma=\pm\tfrac{1}{2}$ denotes spin of the electrons, cf. figure \ref{fig:freeelectronnumbers}.
In this integer space, the number of possible states per unit interval -- the total density of states -- is one. Due to the linear relation between momentum and these integers, the total density of states for free electrons of spin $\sigma$ in momentum space is also constant,
\begin{equation}
\rho_\sigma(p_i) \equiv \frac{1}{L} \frac{1}{p_{i+1}-p_i} = \frac{1}{2\pi}.
\end{equation}

As usual in thermodynamics we will introduce the partition function
\begin{equation}
Z = \sum_n \langle \psi_n | e^{-\beta H} | \psi_n \rangle = e^{-\beta F},
\end{equation}
where $\beta = 1/T$ is the inverse temperature, and $F$ is the free energy. From here you can compute various thermodynamic quantities, especially upon including chemical potentials (in $H$ if you wish). In particular, via various paths familiar from basic statistical mechanics, you can derive the momentum distribution of free fermions in thermal equilibrium
\begin{equation}
\label{eq:FDdist}
\rho_{\mathrm{FD}}(p) = \frac{1}{2\pi} \frac{1}{1+e^{E(p)/T}},
\end{equation}
known as the Fermi-Dirac distribution. Here $E(p)$ is the dispersion relation of the fermions. We will directly compute the full partition function for free fermions in the large volume limit, in a way that will extend to general integrable models where we only have an implicit description of states at asymptotically large volume.

In the large volume limit, states with finite numbers of particles contribute negligibly to the partition function so we will consider the limit $L \rightarrow \infty$ considering states with finite density $N_\sigma/L$, $N_\sigma$ denoting the number of electrons with spin $\sigma$. These $N_\sigma$ particles have distinct momenta that need to occupy $N_\sigma$ of the allowed values of momentum. If a momentum value is taken we will talk of a particle with this momentum, and if it is not, a hole, as in figure \ref{fig:freeelectronnumbers}. Since we want to describe finite density states, let us introduce densities for particles and holes as
\begin{equation*}
\begin{aligned}
L \rho^f_\sigma(p)  \Delta p &= \# \mbox{of particles with spin $\sigma$ and momentum between $p$ and $p+\Delta p$},\\
L \bar{\rho}^f_\sigma(p)  \Delta p &= \# \mbox{of holes with spin $\sigma$ and momentum between $p$ and $p+\Delta p$}.
\end{aligned}
\end{equation*}
By definition these add up to the total momentum density of states, i.e.
\begin{equation}
\label{eq:partholesumconst}
\rho^f_\sigma(p)+\bar{\rho}^f_\sigma(p) = \rho_\sigma(p) = \frac{1}{2\pi}.
\end{equation}
Now, to compute the partition function in a thermodynamic picture we need the free energy $F = E-TS$, in other words the energy and entropy of possible configurations. By definition the energy density of any given state is
\begin{align}
e &= \frac{1}{L} \sum_\sigma \sum_{j=1}^{N_\sigma} E_\sigma (p_j),\\
  &= \sum_j \sum_\sigma E_\sigma (p_j) \frac{p_{j+1} - p_j}{L(p_{j+1} - p_j)},\\
  &= \sum_j \sum_\sigma E_\sigma (p_j) (p_{j+1} - p_j) \rho^f_\sigma (p_j),
\end{align}
where the last line is nicely of the form of a discretized integral, appropriate for the large volume limit. There we get
\begin{equation}
e = \int_{-\infty}^{\infty}dp \sum_\sigma E_\sigma(p) \rho^f_\sigma(p),
\end{equation}
where we write $\rho^f_\sigma(p)$ for the $L\rightarrow \infty$ limit of $\rho^f_\sigma(p_j)$. In a lattice model we would integrate from $0$ to $2\pi$ (given appropriate normalization choices). Next we want to find an expression for the entropy, the logarithm of the number of available states. By definition
\begin{equation}
\Delta S (p_j) = \log \prod_\sigma \frac{(L \Delta p_j \rho_\sigma(p_j))!}{(L \Delta p_j \rho^f_\sigma(p_j))!(L \Delta p_j \bar{\rho}^f_\sigma(p_j))!}
\end{equation}
which in the large volume limit we can approximate via Stirling's formula, $\log n! = n \log n - n + \mathcal{O}(\log n)$, as
\begin{equation}
\Delta S (p_j) = L \Delta p_j \sum_\sigma \rho_\sigma(p_j) \log \rho_\sigma(p_j) - \rho^f_\sigma(p_j) \log \rho^f_\sigma(p_j)- \bar{\rho}^f_\sigma(p_j) \log \bar{\rho}^f_\sigma(p_j).
\end{equation}
In the thermodynamic limit the entropy density is thus given by
\begin{equation}
s = \int_{-\infty}^{\infty} d p \sum_\sigma \rho_\sigma(p) \log \rho_\sigma(p) - \rho^f_\sigma(p) \log \rho^f_\sigma(p)- \bar{\rho}^f_\sigma(p) \log \bar{\rho}^f_\sigma(p).
\end{equation}
Putting all this together we find that the free energy density $f$ at temperature $T$, $f=e-Ts$, is given by
\begin{equation}
f = \int_{-\infty}^{\infty} d p \sum_\sigma E_\sigma(p) \rho^f_\sigma(p) - T(\rho_\sigma(p) \log \rho_\sigma(p) - \rho^f_\sigma(p) \log \rho^f_\sigma(p)- \bar{\rho}^f_\sigma(p) \log \bar{\rho}^f_\sigma(p)).
\end{equation}
This is a functional of the densities $\rho$, and thermodynamic equilibrium corresponds to its stationary point. To find this stationary point we should vary $f$ with respect to $\rho^f_\sigma$ and $\bar{\rho}^f_\sigma$, but these are not independent! The hole and particle densities are constrained by eqn. \eqref{eq:partholesumconst}, which means
\begin{equation}
\delta \bar{\rho}^f_\sigma = - \delta \rho^f_\sigma.
\end{equation}
We then have
\begin{align}
\delta f & = \int_{-\infty}^{\infty} d p \sum_\sigma E_\sigma(p) \delta \rho^f_\sigma(p) - T\left(\log \frac{\rho_\sigma(p)}{\rho^f_\sigma(p)} \delta\rho^f_\sigma(p) + \log \frac{\rho_\sigma(p)}{\bar{\rho}^f_\sigma(p)} \delta\bar{\rho}^f_\sigma(p)\right)\\
& = \int_{-\infty}^{\infty} d p \delta \rho^f_\sigma(p) \left(\sum_\sigma E_\sigma(p) - T \log \frac{\bar{\rho}^f_\sigma(p)}{\rho^f_\sigma(p)}\right) = 0,
\end{align}
from which we conclude
\begin{equation}
\frac{\bar{\rho}^f_\sigma(p)}{\rho^f_\sigma(p)} = e^{E_\sigma(p)/T}.
\end{equation}
Together with eqn. \eqref{eq:partholesumconst} this gives
\begin{equation}
\rho^f_\sigma(p) = \frac{1}{2\pi} \frac{1}{1+e^{E_\sigma(p)/T}},
\end{equation}
which is nothing but the Fermi-Dirac distribution \eqref{eq:FDdist} (here derived in infinite volume). Now we can insert this and the corresponding $\bar{\rho}^f_\sigma$ back into the free energy to find
\begin{equation}
f = - T \int_{-\infty}^{\infty} \frac{dp}{2\pi} \sum_\sigma \log (1+ e^{-E_\sigma(p)/T}).
\end{equation}
This is the well known infinite volume free energy of a Fermi gas.

We would like to follow this approach to describe the thermodynamics of general integrable models, where the relation between particle and hole densities is not as simple as eqn. \eqref{eq:partholesumconst}, but nevertheless known. Let us begin with the integrable model for which this was originally done.

\subsection{The Bose gas}
\label{sec:Bosegas}

The Bose gas, also known as the Lieb-Liniger model, is a system of $N$ bosons interacting via a repulsive delta function interaction. The Hamiltonian is given by
\begin{equation}
H  = - \sum_{j=1}^N \frac{\partial^2}{\partial x_i^2} + 2 c \sum_{i>j} \delta(x_i - x_j),
\end{equation}
with $c>0$, and we consider it on a circle of circumference $L$. This model was `solved' by Bethe ansatz in \cite{Lieb:1963rt}. Based on this the thermodynamics of the model were described by Yang and Yang \cite{Yang:1968rm}, leading to what is now known as the thermodynamic Bethe ansatz. In this section we follow their timeless 1968 paper fairly directly. The nice point about this model is that some things we will have to assume later, can be made precise here. The starting point for our analysis will be the Bethe equations of the Bose gas
\begin{equation}
\label{eq:BosegasBetheeqs}
e^{i p_j L} = \prod_{k\neq j}^N \frac{p_j - p_k + i c}{p_j - p_k - i c},
\end{equation}
from which we see that we have an S-matrix given by
\begin{equation}
S(p_l,p_m)=S(p_l-p_m) = \frac{p_l-p_m-ic}{p_l-p_m+ic}.
\end{equation}
The solutions of these equations are real.\footnote{Consider the equation for the momentum with maximal imaginary part (pick one in case there are multiple), then the right hand side of the equation necessarily has norm greater than or equal to one. The left hand side however has norm less than or equal to one. Therefore we conclude the maximal imaginary part is zero. Similarly, the minimal imaginary part is zero.} The dispersion relation of these bosons is just the free $E(p) = p^2$.

To get the momentum density of states we need to take a logarithm of the Bethe equations, just as we did for free particles above. To do so we note that
\begin{equation}
S(p) = - e^{2 i \arctan{p/c}} \equiv - e^{i\psi(p)},
\end{equation}
so that we get
\begin{equation}
2 \pi I_j = p_j L -i \sum_k \log S(p_j-p_k) = p_j L + \sum_k \left( \psi(p_j-p_k) + \pi \right),
\end{equation}
which is all defined up to the integer $I_j$ defining the branch of the logarithm that we take. In the original paper the factor of $N \pi$ is absorbed in these (then possibly half) integers; we simply take the logarithm of the S-matrix on the right hand side, as this naturally generalizes to any model. These integers $I_j$ are in one to one correspondence with solutions of the Bethe equations, just as for the free particle. To prove this, Yang and Yang introduced what is now known as the Yang-Yang--functional.

\subsubsection{The Yang-Yang--functional}

Let us define
\begin{equation}
B(p_1,\ldots,p_N) = \tfrac{1}{2} L \sum_{l=1}^N p_l^2 - \pi \sum_{j=1}^N (2 I_j + N-1) p_j + \tfrac{1}{2} \sum_{n,m} \left(\psi_1(p_n-p_m)\right),
\end{equation}
where
\begin{equation}
\psi_1(p) = \int_0^p \psi(p^\prime) dp^\prime = \int_0^p 2 \arctan{\tfrac{p^\prime}{c}}\, dp^\prime.
\end{equation}
The nice thing is that by construction $B$ is an `action' with the Bethe equations \eqref{eq:BosegasBetheeqs} as `equations of motion'. Moreover, the matrix $\partial^2 B / \partial k_i \partial k_j$ is positive definite, since the first term in $B$ contributes positively, the second nothing, and the third is positive-semidefinite since $\psi^\prime(p)\geq0$. So $B$ has a unique extremum, a minimum, whose location is determined by solutions to the Bethe equations. Furthermore, all involved quantities clearly depend continuously on $c$ (via the $S$ matrix). Now in the limit $c\rightarrow \infty$ we want to find the wavefunction for $N$ free particles, under the constraint that it vanishes when any two of its arguments coincide, thanks to the infinitely strong repulsion at coincidence. Playing around with this problem a bit in the way that we learn in a course on quantum mechanics, we would find that such wave functions are precisely of Bethe ansatz form, with $S=-1$, precisely the $c \rightarrow \infty$ limit of our S matrix. At this point we have
\begin{equation}
p_j = (2I_j + N-1)\pi/L,
\end{equation}
i.e. the momenta are uniquely identified by the integers $I$ (for a given number of particles $N$). By continuity in $c$ we see that the solutions of the Bethe equations are given by unique sets of distinct momenta in one to one correspondence with sets of distinct integers $I$, which form a complete set of solutions. We can view these $I$'s as quantum numbers for our problem, just as they were for free electrons.

\subsubsection{Thermodynamics}

\begin{figure}%
\centering
\includegraphics[width=8cm]{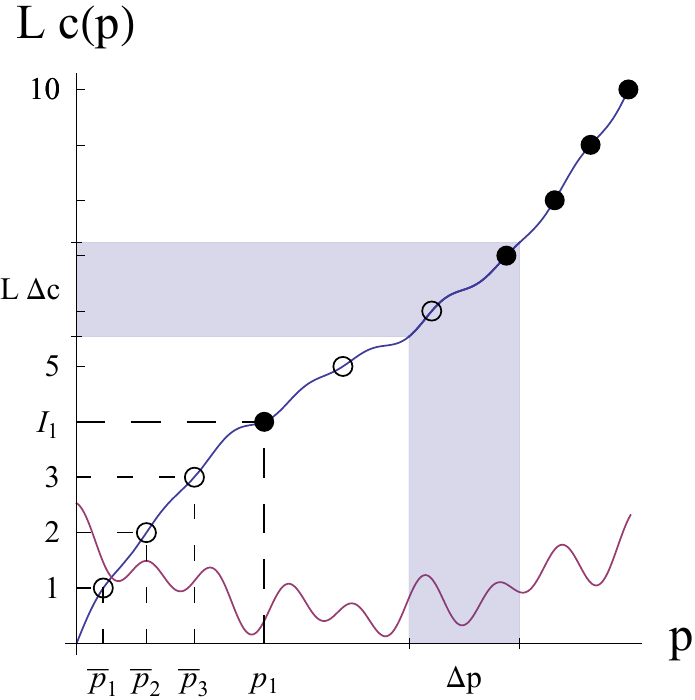}
\caption{The counting function for a hypothetical distribution of roots. The blue line denotes $L$ times the counting function, which takes integer values at fixed values of momenta, indicated along the function by dots. Open dots indicate unoccupied integers (holes), filled dots particles. For instance the first particle momentum $p_1$ corresponds to quantum number $Lc(p_1)=4$. The red line is the (everywhere positive) derivative of the counting function.}
\label{fig:countingfunction}
\end{figure}

Now we are in a position to apply the ideas of the previous section on free fermions to the Bose gas. To start with, we should understand the relation between the quantum numbers and the momenta in more detail. Let us introduce the so-called counting function $c(p)$ as
\begin{equation}
L c(p) = \frac{L}{2\pi} p + \frac{1}{2\pi i} \sum_k \log S(p-p_k).
\end{equation}
For the Bose gas you can explicitly see that this is a monotonically increasing function. Now, if we have a state with quantum numbers $\{I\}$, by definition the particle momenta correspond to the $p$'s for which $L c(p_j) = I_j$. By analogy we then say that any allowed quantum number $J \not\in \{I\}$ represents a hole with momentum $L c(p) = J$. We can schematically depicted this situation in figure \ref{fig:countingfunction}.
The corresponding physical picture is as follows. Since each particle carries energy $p^2$, by monotonicity of the counting function it is clear that the $N$ particle ground state has quantum numbers running between $-\lfloor (N-1)/2\rfloor$ and $\lfloor (N-1)/2\rfloor$ (in unit steps). Excited states now correspond to particles living on the same quantum number lattice (cf. the previous subsection). One or more of them have been moved out of the ground state interval to higher quantum numbers, however, leaving one or multiple `holes' behind in the ground state lattice, cf. figures \ref{fig:freeelectronnumbers} and  \ref{fig:countingfunction}.

As before we introduce densities for the particles and holes as
\begin{equation*}
\begin{aligned}
L \rho^b(p)  \Delta p &= \# \mbox{of particles with momentum between $p$ and $p+\Delta p$},\\
L \bar{\rho}^b(p)  \Delta p &= \# \mbox{of holes with momentum between $p$ and $p+\Delta p$}.
\end{aligned}
\end{equation*}
Again the total density of states in quantum number space is one, which in momentum space picks up a measure factor (Jacobian), cf. figure \ref{fig:countingfunction}, and we find
\begin{equation}
\label{eq:countingfunctiondensityrelationbosegas}
\rho^b(p) + \bar{\rho}^b(p) = \rho(p) = \frac{d c(p)}{dp},
\end{equation}
where we have replaced the discrete derivative by the continuous one appropriate for the thermodynamic limit, and we keep the normalization by $2\pi/L$ introduced when discussing free electrons. In the Bethe equations we encounter sums over particles, which become integrals over densities since as before
\begin{equation*}
\frac{1}{L} \sum_{k\neq j}^N \log S(p_j-p_k) =\sum_{k\neq j}^N \log S(p_j-p_k) \frac{p_k-p_{k+1}}{L(p_k-p_{k+1})} \rightarrow \int_{-\infty}^{\infty} dp^\prime  \log S(p_{(j)}-p^\prime) \rho^b(p^\prime).
\end{equation*}
Using relation \eqref{eq:countingfunctiondensityrelationbosegas} to also express the left hand side of the Bethe equations in terms of densities we find
\begin{equation}
\label{eq:TDLbosegasBetheeqs}
\rho^b(p) + \bar{\rho}^b(p) = \frac{1}{2\pi}  + K \star \rho^b (p),
\end{equation}
where
\begin{equation}
K(p) = \frac{1}{2\pi i} \frac{d}{dp} \log S(p),
\end{equation}
and $\star$ denotes the convolution\footnote{In models where the momenta do not enter the S matrix in difference form, the derivative in $K$ refers to the first argument ($p$ of $S(p,p^\prime)$), while the convolution would become an integral over the second ($p^\prime$). We will only encounter models where we can pick a parametrization that gives a difference form.}
\begin{equation}
f \star g\,(p) \equiv \int_{-\infty}^{\infty} dp^\prime  f(p-p^\prime)g(p^\prime).
\end{equation}
Equation \eqref{eq:TDLbosegasBetheeqs} is the thermodynamic analogue of the Bethe equations, and the analogue of the constraint \eqref{eq:partholesumconst} for free particles (note that eqn. \eqref{eq:TDLbosegasBetheeqs} actually reduces to $\eqref{eq:partholesumconst}$ for a trivial S matrix). Now we are in the same position as we were for free electrons.

The free energy is of the same form as before,
\begin{equation}
f = \int_{-\infty}^{\infty} d p \left(E \rho^b - T\left(\rho \log \rho - \rho^b \log \rho^b- \bar{\rho}^b \log \bar{\rho}^b\right) \right),
\end{equation}
where we recall that for our almost free bosons $E(p) = p^2$. To describe thermodynamic equilibrium we should now vary $f$ with respect to $\rho^b$ and $\bar{\rho}^b$, subject to eqn. \eqref{eq:TDLbosegasBetheeqs} meaning
\begin{equation}
\delta \bar{\rho}^b = - \delta \rho^b  + K \star \delta \rho^b.
\end{equation}
The result is a little more complicated than before
\begin{align}
\delta f & = \int_{-\infty}^{\infty} d p  \left( E\delta \rho^b - T\left(\log \frac{\rho}{\rho^b} \delta\rho^b + \log \frac{\rho}{\bar{\rho}^b} \delta\bar{\rho}^b\right)\right)\\
& = \int_{-\infty}^{\infty} dp  \, \delta \rho^b\left(E - T (\log \frac{\bar{\rho}^b}{\rho^b} + \log\left(1+\frac{\rho^b}{\bar{\rho}^b}\right) \tilde{\star} \,K) \right)
\end{align}
where $\tilde{\star}$ denotes `convolution' from the right,
\begin{equation}
f \, \tilde{\star}\, K(p) = \int_{-\infty}^\infty dp^\prime f(p^\prime)K(p^\prime-p).
\end{equation}
Introducing the pseudo-energy $\epsilon$ by analogy to the free fermion case
\begin{equation}
\frac{\bar{\rho}^b}{\rho^b}(p) = e^{\epsilon(p)/T},
\end{equation}
we see that in thermodynamic equilibrium it needs to satisfy
\begin{equation}
\label{eq:bosegasTBA}
\epsilon(p) = E(p) - T \log (1 + e^{-\epsilon/T}) \,\tilde{\star}\, K
\end{equation}
known as a thermodynamic Bethe ansatz equation. This equation can be numerically solved by iteration, as clearly discussed in appendix A of the original paper \cite{Yang:1968rm}. We briefly discuss some general aspects of solving TBA equations numerically in appendix \ref{app:numTBA}. Given a solution of this equation, the free energy in thermodynamic equilibrium is given by
\begin{equation}
\label{eq:bosegasF}
f = - T \int_{-\infty}^{\infty} \frac{dp}{2\pi} \log (1+  e^{-\epsilon/T}).
\end{equation}
The above formulae are frequently written in terms of a Y function $Y= e^{\epsilon(p)/T}$.

In summary, starting with the Bethe ansatz solution of the one dimensional Bose gas with $\delta$ function interaction, we can continue to use concepts like density of states as we did for free electrons, because individual momenta are still conserved. The nontrivial S matrix of the model now results in an integral equation for the particle density in thermodynamic equilibrium. In this way we reduce the computation of the infinite volume partition function of an \emph{interacting} theory to an integral equation that we can solve rather easily at least numerically, for any value of the coupling $c$.

In a general integrable model the situation is a little more complicated if its excitation spectrum contains bound states of elementary excitations. The XXX spin chain is such a model, and furthermore represents the internals of the chiral Gross-Neveu model.

\subsection{The XXX spin chain}

The Heisenberg XXX spin chain is a one dimensional lattice model with Hamiltonian
\begin{equation}
\label{eq:XXXham}
H = - \frac{J}{4} \sum_{i=1}^{\NF} \left(\vec{\sigma}_i \cdot \vec{\sigma}_{i+1}-1\right) ,
\end{equation}
where $\vec{\sigma}$ is the vector of Pauli matrices. We take the lattice to be periodic; $\sigma_{\NF+1}=\sigma_{1}$. This Hamiltonian acts on a Hilbert space given by $\NF$ copies of $\mathbb{C}^2$, one for each lattice site $i$. Identifying $(1,0)$ as $|\hspace{-2pt} \uparrow \hspace{1pt} \rangle$  and $(0,1)$ as $|\hspace{-2pt} \downarrow \hspace{1pt} \rangle$ , states in this Hilbert space can be viewed as chains of spins, in this case closed. For $J>0$ this is a model of a ferromagnet where spins prefer to align, while for $J<0$ we have an antiferromagnet where spins prefer to alternate.

The Bethe equations for this model are
\begin{equation}
\label{eq:homogeneousXXXbetheequations}
e^{i \scp_i \NF} \prod_{j=1}^\NA S^{11}(v_i-v_j) = -1,
\end{equation}
where
\begin{equation}
\label{eq:XXXmomentum}
\scp_i = \scp(v_i), \hspace{20pt} \scp(v) = -i \log S^{1f}(v),
\end{equation}
and
\begin{equation}
\label{eq:cGNSmatdef}
S^{11}(w) = \frac{w-2i}{w+2i} , \, \, \, S^{1f}(w) = \frac{w+i}{w-i} .
\end{equation}
These equations are the homogeneous limit of the auxiliary Bethe equations of the chiral Gross-Neveu model we will encounter later, where the ``$f$'' will stand for the fermions of this model. The reason for the remaining notation will become apparent soon. The energy eigenvalue associated to a solution of these Bethe equations is
\begin{equation}
\scE = \sum_i \scE_1(v_i), \,\, \mbox{ where }\,\, \scE_1(v) = -2 J \frac{1}{v^2 +1}.
\end{equation}

\subsubsection{The string hypothesis}

\begin{figure}%
\centering
\includegraphics[width=8cm]{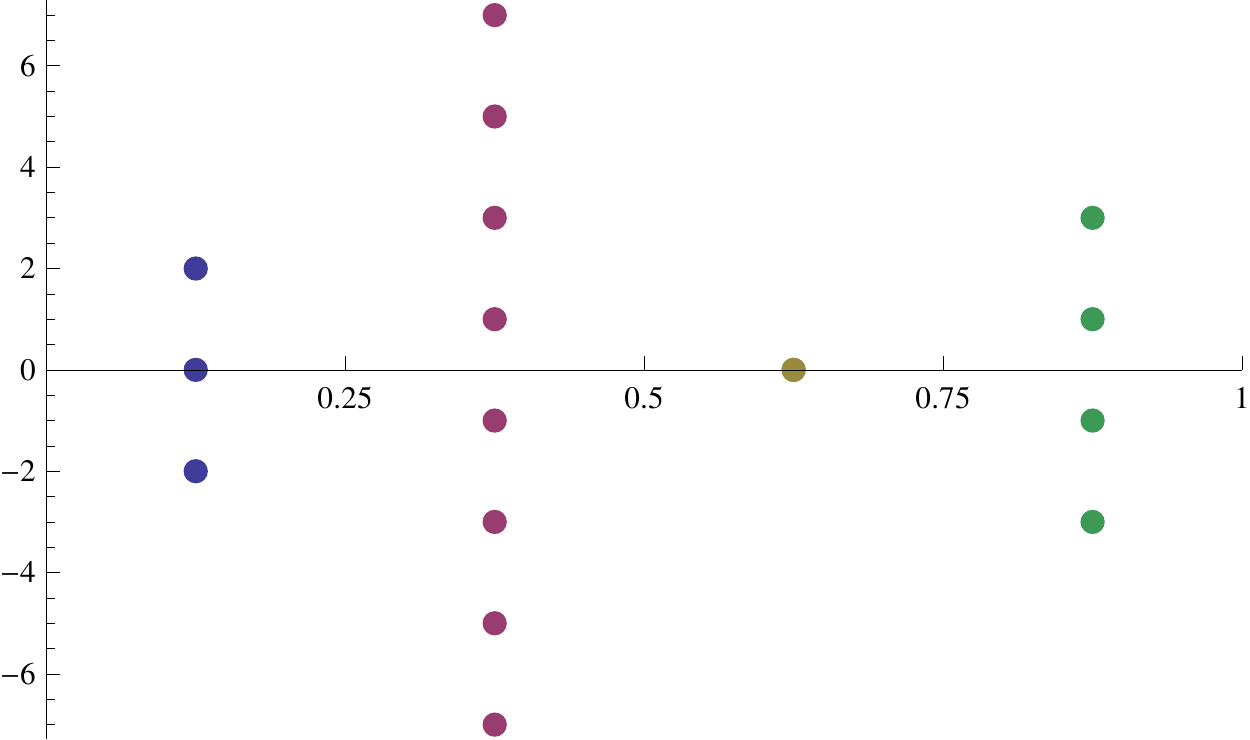}
\caption{Bethe strings. Bethe strings are patterns of rapidities with spacing $2i$. Here we illustrate strings of length three, eight, one and four, with center $1/8$, $3/8$, $5/8$ and $7/8$ respectively.}
\label{fig:bethestrings}
\end{figure}

To describe the thermodynamics of this model, we would like to understand the type of solutions these equations can have, specifically as we take the system size $\NF$ to infinity.\footnote{Here we directly follow the discussion of this topic in \cite{vanTongeren:2013gva}.} The situation will be considerably different from the Bose gas that we just discussed, because here we can have solutions with complex momenta,\footnote{They exist for instance for the Bethe equations with $\NF=5$, $\NA=2$.} For real momenta nothing particular happens in our equations, and we simply get many more possible solutions as $\NF$ grows. If we consider a solution with complex momenta, however, say a state with $\mbox{Im}(\scp_1)>0$, we have an immediate problem:
\begin{equation}
e^{i \scp_q \NF} \rightarrow 0 , \, \, \,\,\,\, \mbox{as} \,\,\, \NF \rightarrow \infty.
\end{equation}
We see that the only way a solution containing $p_1$ can exist in this limit is if this zero is compensated by a pole in one of the $S^{11}$ (eqn. \eqref{eq:cGNSmatdef}), which can be achieved by setting
\begin{equation}
v_2 = v_1 + 2 i.
\end{equation}
At this point we have fixed up the equation for $\scp_1$, but we have introduced potential problems in the equation for $\scp_2$. Whether there is a problem can be determined by multiplying the equations for $\scp_1$ and $\scp_2$ so that the singular contributions of their relative S-matrix cancel out
\begin{equation*}
e^{i (\scp_1+\scp_2) \NF} \prod_{i\neq1}^\NA S^{11}(v_1-v_i) \prod_{i\neq2}^\NA S^{11}(v_2-v_i)  = e^{i (\scp_1+\scp_2) \NF} \prod_{i\neq1,2}^\NA S^{11}(v_1-v_i) S^{11}(v_2-v_i) = 1,
\end{equation*}
and the two particles together effectively scatter with the others by the S matrix
\begin{equation*}
S^{21}(v-v_i) = S^{11}(v_1-v_i) S^{11}(v_2-v_i) = \frac{v-v_i-3i}{v-v_i+3i} \frac{v-v_i-i}{v-v_i+i},
\end{equation*}
where $v = (v_1+v_2)/2 = v_1 + i$. If the sum of their momenta is real this equation is fine, and the momenta can be part of a solution to the Bethe equations. In terms of rapidities this solution would look like
\begin{equation}
\label{eq:XXXtwostring}
v_1 = v - i , \, \, \, v_2 = v + i , \, \, \, v \in \mathbb{R}.
\end{equation}
On the other hand, if the sum of our momenta has positive imaginary part we are still in trouble.\footnote{By rearranging the order of our argument (the particles considered) we do not have to consider the case where the remaining imaginary part is of different sign.} In this case, since we should avoid coincident rapidities in the Bethe ansatz, the only way to fix things is to have a third particle in the solution, with rapidity
\begin{equation}
v_3 = v_2 + 2 i.
\end{equation}
As before, if now the total momentum is real the equations are consistent and these three rapidities can form part of a solution. If not, we continue this process and create a bigger configuration, or run off to infinity. These configurations in the complex rapidity plane are known as Bethe strings, illustrated in figure \ref{fig:bethestrings}.
Since our spin chain momentum $\scp$ has positive imaginary part in the lower half of the complex rapidity plane and vice versa, strings of any size can be generated in this fashion by starting appropriately far below the real line.\footnote{In other models the pattern of possible string configurations can be quite complicated, see e.g. chapter 9 of \cite{Takahashi:book} for the XXZ spin chain as a classic example, or \cite{Saleur:2000bq} and \cite{Arutyunov:2012zt} for more involved examples.} Concretely, a Bethe string with $Q$ constituents and rapidity $v$ is given by the configuration
\begin{equation}
\label{eq:XXXstringconfiguration}
\{v_Q\} \equiv \{v-(Q+1-2j)i|j=1,\ldots,Q\},
\end{equation}
where $v \in \mathbb{R}$ is called the center of the string. Full solutions of the Bethe equation in the limit $\NF\rightarrow \infty$ can be built out of these string configurations. Let us emphasize that these string solutions only ``exist'' for $\NF\rightarrow \infty$. At large but finite $\NF$ root configurations are typically only of approprimate string form.

These (Bethe) strings can be interpreted as bound states, having less energy than sets of individual real magnons.\footnote{The corresponding Bethe wave-function also shows an exponential decay in the separation of string constituents.} For example, the energy of the two-string \eqref{eq:XXXtwostring} is
\begin{equation}
\scE_2(v) = \scE(v_1)+\scE(v_2) = -2 J \left( \frac{1}{(v-i)^2+1}+\frac{1}{(v+i)^2+1}\right) = -2 J \frac{2}{v^2+2^2},
\end{equation}
which is less than that of any two-particle state with real momenta:
\begin{equation}
\scE_2(v)<\scE(\tilde{v}_1)+\scE(\tilde{v}_2) \, \, \,\mbox{ for }\, v,\tilde{v}_{1,2} \in \mathbb{R} \, \,\,\mbox{ (real momenta).}
\end{equation}
Similarly, the energy of a $Q$-string is lower than that of $Q$ separate real particles and is given by
\begin{equation}
\scE_Q(v) = \sum_{v_j \in \{v_Q\}} \scE(v_j) = -2 J \frac{Q}{v^2+Q^2}.
\end{equation}
This is most easily shown by noting that
\begin{equation}
\scE(v) = J \frac{d\scp(v)}{dv},
\end{equation}
and the particularly simple expression for the momentum of a $Q$-string
\begin{equation}
\label{eq:fusedmomentum}
\scp^Q(v) = i \log \frac{v-Q i}{v+Q i},
\end{equation}
as follows by cancelling numerators and denominators in the product $\frac{v_1-i}{\cancel{v_1+i}}\frac{\cancel{v_2-i}}{v_2+i}\ldots \frac{v_Q-i}{v_Q+i}$ as indicated.

We have just determined that the possible solutions of the Bethe equations in the limit $\NF\rightarrow \infty$ are built out of elementary objects called Bethe strings (a one-string being a normal magnon). Interpreting them as bound states, the spectrum thus obtained is reflected by an appropriate pole in the two-particle S-matrix. This example is not a field theory, but such patterns generically hold there (as well).

So far so good, but ultimately we are interested in thermodynamic limits, meaning we should take $\NF\rightarrow \infty$ with $\NA/\NF\leq1/2$ fixed -- the number of magnons goes to infinity as well. In this limit the analysis above is no longer even remotely rigorous since an ever growing product of magnon S-matrices with complex momenta can mimic the role of the pole in our story for example. Still, since such solutions seem rather atypical, and at least low magnon density solutions should essentially conform to the string picture, we can hypothesize that `most' of the possible solutions are made up of string complexes, in the sense that they are the ones that give measurable contributions to the free energy. Indeed in the XXX spin chain there are examples of solutions that do not approach string complexes in the thermodynamic limit \cite{Woynarovich:1981ca,Woynarovich:1982,Babelon:1982mc}, but nonetheless the free energy is captured correctly by taking only string configurations into account \cite{Tsvelick:1983}. The assumption that all thermodynamically relevant solutions to the Bethe equations are built up out of such string configurations, and which form these configurations take, goes under the name of the \emph{string hypothesis}. More details and references on the string hypothesis can for example be found in chapter four of \cite{Korepin}.

\subsubsection*{Bethe equations for string configurations}

With our string hypothesis for possible solutions in the thermodynamic limit, we would like to group terms in the Bethe equations accordingly -- the $N_a$ magnons of a given solution of the Bethe equations should arrange themselves into combinations of string complexes. Denoting the number of bound states of length $Q$ occurring in a given configuration by $N_Q$ we have
\begin{equation}
\prod_{j=1}^{\NA} \rightarrow \prod_{Q=1}^{\infty} \prod_{l=1}^{N_{Q}} \prod_{j \in \{v_{Q,l}\}},
\end{equation}
under the constraint
\begin{equation}
\sum_{Q=1}^\infty Q N_Q = \NA.
\end{equation}
We can then appropriately represent the Bethe equations as
\begin{align}
e^{i \scp_i \NF} \prod_{Q=1}^{\infty} \prod_{l=1}^{N_{Q}} S^{1Q} (v_i-v_{Q,l}) & = -1,
\end{align}
where
\begin{align}
&S^{1 Q}(v-w_Q) \equiv \prod_{w_j \in \{w_Q\}} S^{1 1}(v - w_j).
\end{align}
At this point not all $\NA$ Bethe equations are independent anymore, as some magnons are bound in strings -- only their centers matter. We already saw that we can get the Bethe equation for the center of a bound state by taking a product over the Bethe equations of its constituents, so that our (complete) set of Bethe equations becomes
\begin{align}
\label{eq:XXXBetheeqsstrings}
e^{i \scp^P_r \NF} \prod_{Q=1}^{\infty} \prod_{l=1}^{N_{Q}} S^{PQ} (v_{P,r}-v_{Q,l}) & = (-1)^P,
\end{align}
where
\begin{align}
&S^{PM}(v_P-w) \equiv \prod_{v_i \in \{v_P\}} S^{1M}(v_i-w) .
\end{align}
Note that we include the term with $(Q,l)=(P,r)$ in the product above since we took the product in the Bethe equations \eqref{eq:homogeneousXXXbetheequations} to run over all particles. Since $S^{PP}(0) = (-1)^{P^2} = (-1)^P$ however, we could cancel this $(Q,l)=(P,r)$ term against the $(-1)^P$ in the Bethe equations for string configurations if we wanted to.

Physically these expressions represent the scattering amplitudes between the particles indicated by superscripts. These products of constituent S-matrices typically simplify, but their concrete expressions are not important for our considerations (yet); what is important is that they exist and only depend on the centers of the strings, i.e. the overal momenta of the bound state configurations. Combining a set of magnons into a string (bound state) is known as fusion, and the above product denotes the fusion of the corresponding scattering amplitude. You might have encountered similar ideas applied to obtain bound state S-matrices from fundamental ones for instance in \cite{Bombardelli:2016scq}, here we just did it at the diagonalized level.

\subsubsection{Thermodynamics}

We now have a grasp on the types of solutions of our Bethe equations in the thermodynamic limit, though this is far from rigorous. We will assume that our classification of possible solutions in terms of strings accurately describes the system in the thermodynamic limit. With this assumption we can proceed as before and derive the thermodynamic Bethe ansatz equations.

We begin with the Bethe equations in logarithmic form, introducing an integer $I$ in each equation which labels the possible solutions
\begin{align}
- 2 \pi I^P_r &=  \NF p^{P}(v_{P,r}) -i \prod_{Q=1}^{\infty} \prod_{\substack{l=1\\ l \neq r}}^{N_{Q}} \log{S^{PQ}(v_{P,r}-v_{Q,l})} .
\end{align}
We choose to define the integer with a minus sign for reasons we will explain shortly. As by now usual, the solutions to these equations become dense
\begin{equation}
\,\,\,\, v_i-v_j \sim \mathcal{O}(1/\NF),
\end{equation}
and we generalize the integers $I$ to counting functions of the relevant rapidity (momentum). Concretely
\begin{align}
\NF c^P (u) &= - \NF \frac{p^{P}(u)}{2\pi} - \frac{1}{2\pi i} \sum_{Q=1}^{\infty} \sum_{\substack{l=1\\ l \neq r}}^{N_{Q}} \log{S^{PQ}(u-v_{Q,l})} ,
\end{align}
so that
\begin{equation}
\NF c^P(v_l) = I^P_l.
\end{equation}
Importantly, in this case we \emph{assume} that the counting functions are monotonically increasing functions of $u$ provided their leading terms are,\footnote{Here we do not have a convenient positive definite Yang-Yang functional at our disposal. It is not obvious how to prove that these functions are monotonically increasing for given excitation numbers without knowing the precise root distribution, which is what we are actually trying to determine. We may consider it part of the string hypothesis by saying we are not making a mistake in treating the thermodynamic limit as the \emph{ordered} limits $\NF\rightarrow\infty$, then $\NA\rightarrow \infty$, in which case the statement does clearly hold. A discussion with similar statements can be found on the first page of section six in \cite{Faddeev:1996iy}.} and here indeed we have
\begin{equation}
\frac{1}{2\pi} \frac{d \scp^P(v)}{dv}<0,
\end{equation}
the reason for our sign choice above. Clearly in general we have
\begin{equation}
c(w_i)-c(w_j) = \frac{I_i-I_j}{\NF}.
\end{equation}
Introducing particle and hole densities as before, except now in rapidity space, we get
\begin{equation}
\rho^P(v) + \bar{\rho}^P(v) = \frac{d c^P (v)}{dv},
\end{equation}
and explicitly taking the derivative of the counting functions gives us the thermodynamic analogue of the Bethe-Yang equations as
\begin{equation}
\label{eq:XXXdensityeqn}
\rho^P(v) + \bar{\rho}^P(v) = -\frac{1}{2\pi} \frac{d \scp^P(v)}{dv} - K^{PQ} \star \rho^Q(v),
\end{equation}
where we implicitly sum over repeated indices, and defined the kernels $K$ as the logarithmic derivatives of the associated scattering amplitudes
\begin{equation}
\label{eq:kerneldef}
K^\chi(u) = \pm \frac{1}{2\pi i}\frac{d}{du} \log S^\chi (u),
\end{equation}
where $\chi$ denotes an arbitrary set of particle labels. The sign is chosen such that the kernels are positive, in this case requiring minus signs for the $K^{M}$.\footnote{Unfortunately we cannot define a notation which uniformizes both the Bethe-Yang equations in the way we did and automatically gives positive kernels.} As before the Bethe-Yang equations come in by giving us the hole densities as functions of the particle densities. Varying eqs. \eqref{eq:XXXdensityeqn} gives
\begin{align}
\delta \rho^P + \delta \bar{\rho}^P &=  - K^{PQ}\star \delta \rho_Q,
\end{align}
Writing this schematically as\footnote{Apologies for the immediate mismatch of signs, but this is the general form we would like to take, and considering eqn. \eqref{eq:TDLbosegasBetheeqs} there is clearly no uniform sign choice.}
\begin{equation}
\label{eq:vardensityeqgen}
\delta \rho^i + \delta \bar{\rho}^i = K^{ij}\star \delta \rho^j,
\end{equation}
after a little algebra we get the variation of the entropy
\begin{equation}
\frac{\delta s}{\delta\rho^j(u)} = \log \frac{\bar{\rho}^j}{\rho^j}(u) + \log \left(1 + \frac{\rho^i}{\bar{\rho}^i}\right) \tilde{\star}\, K^{ij} (u),
\end{equation}
where again $\tilde{\star}$ denotes `convolution' from the right (now in $u$). The variation of the other terms is immediate, and $\delta F=0$ results in the \emph{thermodynamic Bethe ansatz equations}
\begin{equation}
\log \frac{\bar{\rho}^j}{\rho^j} =\frac{E_j}{T} - \log \left(1 + \frac{\rho^i}{\bar{\rho}^i}\right)\star K^{ij},
\end{equation}
where by conventional abuse of notation we dropped the tilde on the `convolution'. We will henceforth denote the combination $\frac{\bar{\rho}^j}{\rho^j}$ by the \emph{Y functions} $Y_j$, meaning the TBA equations read
\begin{equation}
\label{eq:TBAgen}
\log Y_j =\frac{E_j}{T} - \log \left(1 + \frac{1}{Y_i}\right)\star K^{ij}.
\end{equation}
Taking into account the generalized form of eqs. \eqref{eq:XXXdensityeqn} as
\begin{equation}
 \rho^i + \bar{\rho}^i = \frac{1}{2\pi}\frac{dp^i}{du} +  K^{ij}\star \rho^j,
\end{equation}
on a solution of the TBA equations the free energy density is given by
\begin{equation}
f = -T\int_{-\infty}^\infty du \frac{1}{2\pi} \frac{dp_j}{du} \log\left(1+\frac{1}{Y_j}\right).
\end{equation}

Specifying our schematic notation to eqs. \eqref{eq:XXXdensityeqn} gives
\begin{align}
\log Y_P &= \frac{ \scE_P}{T} + \log \left(1 + \frac{1}{Y_Q}\right)\star K^{QP},
\end{align}
and
\begin{equation}
f =T\sum_P \int_{-\infty}^\infty du \frac{1}{2\pi} \frac{d\scp_P}{du} \log\left(1+\frac{1}{Y_P}\right).
\end{equation}
Note the changes of signs due to our conventions on $K$ and $\scp$ compared to eqs. \eqref{eq:bosegasTBA} and \eqref{eq:bosegasF}. In stark contrast to the Bose gas, here we are dealing with an infinite set of equations for infinitely many functions, all functions appearing in each equation.

At this point the generalization to an arbitrary model is hopefully almost obvious, with the exception of the string hypothesis which depends on careful analysis of the Bethe(-Yang) equations for a particular model. If we have this however, we can readily determine the complete set of Bethe(-Yang) equations analogous to the procedure to arrive at eqs. \eqref{eq:XXXBetheeqsstrings}. From there we immediately get the analogue of eqs. \eqref{eq:XXXdensityeqn} by a logarithmic derivative. Note that since we like to think of densities as positive we may have to invert the Bethe(-Yang) equations for a specific particle type to make sure the counting function is defined to be monotonically increasing, just like we did above. This is all we need to specify the general TBA equations \eqref{eq:TBAgen} to a given model. Let us quickly do this for our main field theory example of the chiral Gross-Neveu model.

\subsection{The chiral Gross-Neveu model}

The $\mbox{SU}(N)$ chiral Gross-Neveu model is a model of $N$ interacting Dirac fermions with Lagrangian\footnote{Our $\gamma$ matrices are defined as $\gamma_0 = \sigma_1$, $\gamma_1 = i \sigma_2$, $\gamma_5 = \gamma_0 \gamma_1$, where $\gamma_{0,1}$ form the Clifford algebra $\{ \gamma_\mu, \gamma_\nu \} = 2 \eta^{\mu \nu}$ with $\eta = \mbox{diag}(1,-1)$. Note that $\gamma_5$ is Hermitian. As usual $\bar{\psi} = \psi^\dagger \gamma_0$ and $\slashed{\partial} = \gamma^\mu \partial_\mu$.}
\begin{equation}
\mathcal{L}_{cGN} = \bar{\psi}_a i\slashed{\partial} \psi^a + \frac{1}{2} g_s^2\left( (\bar{\psi}_a \psi^a)^2- (\bar{\psi}_a \gamma_5 \psi^a)^2\right) - \frac{1}{2} g_v^2 (\bar{\psi}_a \gamma_\mu \psi^a)^2,
\end{equation}
where $a=1,\ldots,N$ labels the $N$ Dirac spinors. This Lagrangian has $\mbox{U}(N) \times \mbox{U}(1)_c$ symmetry, where viewed as an $N$-component vector the spinors transform in the fundamental representation of $\mbox{U}(N)$, and $\mbox{U}(1)_c$ denotes the chiral symmetry $\psi \rightarrow e^{i \theta \gamma_5} \psi$. The full spectrum of this theory contains $N-1$ $\mbox{SU}(N)$ multiplets of interacting massive fermions, and massless excitations which carry this chiral $\mbox{U}(1)$ charge that decouple completely.\footnote{These facts are far from obvious looking at the Lagrangian, see e.g. section 2.4.1 in \cite{vanTongeren:2013gva} for a brief discussion with references. Because of the decoupling of the $U(1)$ mode, $g_v$ is typically put to zero in the chiral Gross-Neveu Lagrangian. Keeping $g_v\neq0$, however, is useful in demonstrating equivalence to the $\mathrm{SU}(N)$ Thirring model.} We will focus on the $\mathrm{SU}(2)$ model.

As a relativistic model the dispersion relation of the fermions is
\begin{equation}
E^2 - p^2 = m^2,
\end{equation}
where $m$ is the mass of the fermions. It will be convenient to parametrize energy and momenta in terms of a rapidity $u$ as\footnote{We choose this unconventional normalization of $u$ to get Bethe-Yang equations in `the simplest' form. The relation to the rapidity of D. Bombardelli's article \cite{Bombardelli:2016scq} is simply $\theta = \pi u /2$.}
\begin{equation}
E = m \cosh{\tfrac{\pi u_i}{2}}, \, \, \,
p = m \sinh{\tfrac{\pi u_i}{2}}.
\end{equation}
Note that Lorentz boosts act additively on the rapidity, and therefore by Lorentz invariance the two-body S-matrix is a function of the difference of the particles' rapidities only.

The spectrum of the $\mathrm{SU}(2)$ chiral Gross-Neveu model contains two species of fermions corresponding to $\mathrm{SU}(2)$ spin up and down. This model can be ``solved'' in the spirit of factorized scattering \cite{Zamolodchikov:1978xm}, as discussed for instance in the article by D. Bombardelli \cite{Bombardelli:2016scq}. For the $\mathrm{SU}(2)$ chiral Gross-Neveu model the upshot is that the scattering of two fermions of equal spin has amplitude
\begin{equation}
\, S^{ff}(u) = -\frac{\Gamma(1-\frac{u}{4i})\Gamma(\frac{1}{2}+\frac{u}{4i})}{\Gamma(1+\frac{u}{4i})\Gamma(\frac{1}{2}-\frac{u}{4i})}.
\end{equation}
The relative scattering of fermions with opposite spin is fixed by $\mathrm{SU}(2)$ invariance, which leads to a matrix structure matching the $R$ matrix of the XXX spin chain. Diagonalizing the associated transfer matrix results in the Bethe-Yang equations
\begin{align}
\label{eq:cGNfullBethe1}
&e^{i p_j L} \prod_{m=1}^\NF S^{ff} (u_j-u_m)\prod_{i=1}^\NA S^{f1}(u_j-v_i) = -1,\\
\label{eq:cGNfullBethe2}
&\prod_{m=1}^\NF S^{1f}(v_i-u_m)\prod_{j=1}^\NA S^{11}(v_i-v_j)= -1.
\end{align}
which apply in an asymptotically large volume limit, suiting us just fine in the thermodynamic limit. The amplitudes $S^{11}$, $S^{1f}$ and $S^{f1}(v) = S^{1f}(v)$ are as defined in the previous section in equation \eqref{eq:cGNSmatdef}. The $N_a$ auxiliary excitations with rapidities $v_j$ correspond to changing the $\mathrm{SU}(2)$ spin fermions from up to down; the ``vacuum'' of the transfer matrix was made up of spin up fermions (cf. spin up states in the XXX spin chain). Note that the equations for the auxiliary excitations become the XXX Bethe equations of the previous section in the limit $u_m\rightarrow0$.

\subsubsection*{String hypothesis}

The two types of fermions of the chiral Gross-Neveu model do not form physical bound states -- there is no appropriate pole in the S matrix.\footnote{In our conventions, bound states must have $\mbox{Im}(u) \in (0,2i)$, see e.g. \cite{Dorey:1996gd} or section 2.4.1 of \cite{vanTongeren:2013gva}.} However, to take a thermodynamic limit we need to consider finite density states, meaning we will be taking the limit $L \rightarrow \infty$, but also $\NF\rightarrow \infty$ and $\NA \rightarrow \infty$ keeping $\NF/L$ and $\NA/\NF$ fixed and finite. At the auxiliary level we are hence taking the infinite length limit of our XXX spin chain, where we did encounter bound states. The analysis leading to these string solutions is not affected by including the real inhomogeneities $u_m$ corresponding to the physical fermions of the chiral Gross-Neveu model. The only difference is that here the XXX magnons are auxiliary excitations, meaning they carry no physical energy or momentum, and hence the Bethe string solutions lose their interpretation as physical bound states. Nothing changes with regard to them solving the Bethe-Yang equations in the thermodynamic limit however, and we need to take them into account. For the $\mathrm{SU}(2)$ chiral Gross-Neveu model we will hence make the string hypothesis that the solutions of its Bethe-Yang equations are given by
\begin{itemize}
\item{Fermions with real momenta}
\item{Strings of auxiliary magnons of any length with real center}
\end{itemize}
Fusing the Bethe-Yang equations \eqref{eq:cGNfullBethe1} and  \eqref{eq:cGNfullBethe2} then gives
\begin{align}
\label{eq:cGNBethestrings1}
e^{i p_j L} \prod_{m\neq j}^\NF S^{ff} (u_j-u_m) \prod_{Q=1}^{\infty} \prod_{l=1}^{N_{Q}} S^{fQ} (u_j-v_{Q,l}) & = -1,\\
\label{eq:cGNBethestrings2}
\prod_{m=1}^\NF S^{Pf}(v_{P,r}-u_m)\prod_{Q=1}^{\infty} \prod_{l=1}^{N_{Q}} S^{PQ} (v_{P,r}-v_{Q,l}) & = (-1)^P,
\end{align}
where
\begin{align}
&S^{\chi Q}(v-w_Q) \equiv \prod_{w_j \in \{w_Q\}} S^{\chi 1}(v - w_j) , \hspace{20pt} \chi = f,1,
\end{align}
and
\begin{align}
&S^{P\chi}(v_P-w) \equiv \prod_{v_i \in \{v_P\}} S^{1\chi}(v_i-w) , \hspace{20pt} \chi = f,Q.
\end{align}

\subsubsection*{Thermodynamics}

Via the counting functions we get the thermodynamic analogue of the Bethe-Yang equations
\begin{align}
\label{eq:cGNdensityeqn1}
\rho^f(u) + \bar{\rho}^f(u) &= \frac{1}{2\pi} \frac{dp(u)}{du} + K^{ff} \star \rho^f(u) - K^{fQ} \star \rho_Q(u) ,\\
\label{eq:cGNdensityeqn2}
\rho^P(v) + \bar{\rho}^P(v) &= K^{Pf} \star \rho^f(u) - K^{PQ} \star \rho^Q(u),
\end{align}
where again the kernels are defined as in eqn. \eqref{eq:kerneldef}, positivity of the kernels here requiring minus signs for $K^{fP}$ and $K^{Mf}$. From our general result above we then find the TBA equations
\begin{align}
\label{eq:cGNcanonicalTBA1}
\log Y_f &= \frac{E}{T} - \log \left(1 + \frac{1}{Y_f}\right)\star K^{ff} - \log \left(1 + \frac{1}{Y_Q}\right)\star K^{Qf},\\
\label{eq:cGNcanonicalTBA2}
\log Y_P &= \log \left(1 + \frac{1}{Y_Q}\right)\star K^{QP} + \log \left(1 + \frac{1}{Y_f}\right)\star K^{fP},
\end{align}
and free energy density
\begin{equation}
f = - T \int_{-\infty}^\infty du \frac{1}{2\pi} \frac{dp}{du} \log\left(1+\frac{1}{Y_f}\right).
\end{equation}

The thermodynamics of the chiral Gross-Neveu model (and the XXX spin chain), are determined through an infinite number of integral equations, each directly coupled to all others. Fortunately, this structure can be simplified.

\subsection{From TBA to Y system}
\label{sec:TBAtoY}

\begin{figure}%
\centering
\includegraphics[width=6cm]{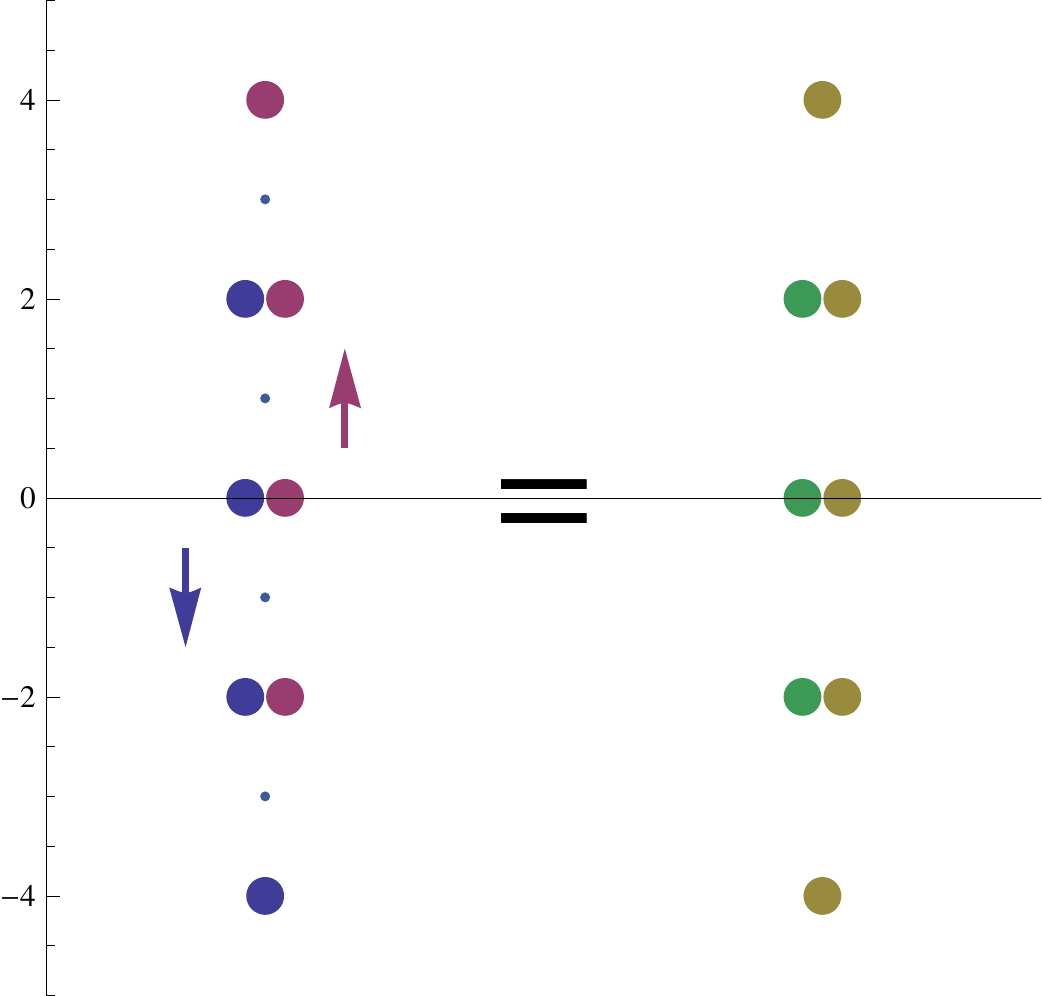}
\caption{The discrete Laplace equation for strings. Shifting a length $Q$ string configuration up by $i$ and another down by $i$ gives a configuration equivalent to two unshifted strings, one of length $Q+1$ and another of length $Q-1$, here illustrated for $Q=4$. The small dots indicate the position of the rapidities before shifting.}
\label{fig:stringdiscretelaplace}
\end{figure}

In problems where there are (auxiliary) bound states the TBA equations can typically be rewritten in a simpler fashion. This is possible for the intuitive reason illustrated in figure \ref{fig:stringdiscretelaplace}.
Since we obtained all bound state S matrices by fusing over constituents, provided $S$ has no branch cuts the figure shows that
\begin{equation}
\label{eq:Sdiscretelaplace}
\frac{S^{\chi Q+1}(v,u)S^{\chi Q-1}(v,u)}{S^{\chi Q}(v,u+i)S^{\chi Q}(v,u-i)} = 1,
\end{equation}
where $\chi $ is any particle type and we have reinstated a dependence on two arguments for clarity. We see that (the logs of) our S-matrices satisfy a discrete Laplace equation. Hence the associated kernels would naively satisfy
\begin{equation}
\label{eq:naivekernelidentity}
K^{\chi Q}(v,u+i) + K^{\chi Q}(v,u-i) - (K^{\chi Q+1}(v,u) + K^{\chi Q-1}(v,u)) = 0.
\end{equation}
However, when we shift $u$ by $\pm i$ we may generate a pole in $K(v,u+i)$ for some real value of $v$. This can lead to a discontinuity in integrals involving K such as those in the TBA equations. Therefore we need to understand what exactly we mean by this equation. To do so, let us introduce the kernel $s$
\begin{equation}
s(u)=\frac{1}{4\cosh\frac{\pi u}{2}},
\end{equation}
and the operator $s^{-1}$ that in hindsight will properly implement our shifts
\begin{equation}
f \star s^{-1} (u) = \lim_{\epsilon\rightarrow 0} \left(f(u+i-i\epsilon) + f(u-i+i\epsilon)\right),
\end{equation}
which satisfy
\begin{equation}
s\star s^{-1}(u) = \delta(u).
\end{equation}
Note that $s^{-1}$ has a large null space, so that $f \star s^{-1} \star s \neq f$ in general; we will see examples of this soon. This kernel can now be used to define
\begin{equation}
\label{eq:Kplus1invdef}
(K+1)^{-1}_{PQ} = \delta_{P,Q} - I_{PQ} s,
\end{equation}
where the incidence matrix $I_{PQ} = \delta_{P,Q+1} +  \delta_{P,Q-1}$, and $\delta_{M,N}$ is the Kronecker delta symbol. This is defined so that
\begin{equation}
(K+1)_{MP}\star (K+1)^{-1}_{PN} = 1_{M,N},
\end{equation}
where $1$ denotes the identity in function and index space: $1_{M,N}=\delta(u) \delta_{M,N}$. In other words, the kernel $K^{PQ}$ introduced above is supposed to satisfy
\begin{equation}
\label{eq:KQMidentity}
K^{PQ} - (K^{PQ+1}+K^{PQ-1} )\star s = s \, I_{PQ},
\end{equation}
which we can prove by Fourier transformation, see appendix \ref{app:kernelidentities} for details. Similarly we have
\begin{equation}
K^{fQ} - (K^{fQ+1}+K^{fQ-1} )\star s = s \, \delta_{Q1}.
\end{equation}
Note how the naive picture of eqn. \eqref{eq:naivekernelidentity} misses the right hand side of these identities. If a set of TBA equations contains other types of kernels these typically also reduce to something nice after acting with $(K+1)^{-1}$.

\subsubsection*{Simplified TBA equations}

With these identities we can rewrite the auxiliary TBA equations \eqref{eq:cGNcanonicalTBA2} for the chiral Gross-Neveu model as
\begin{equation}
\log Y_Q = \log (1 + Y_{Q+1})(1 + Y_{Q-1}) \star s+ \delta_{Q,1}\log \left(1 + \frac{1}{Y_f}\right) \star s.
\end{equation}
This follows from convoluting the equations for $Y_{Q\pm1}$ with $s$ and subtracting them from the equation for $Y_Q$. Note the remarkable simplification that all infinite sums have disappeared! These TBA equations are not surprisingly known as simplified TBA equations, versus the canonical ones we derived them from.

We should be careful not to oversimplify however. The fact is that $(K+1)^{-1}$ has a null space that is typically of physical relevance. For example, if we take our chiral Gross-Neveu model and turn on a (constant) external magnetic field $B$ coupling to the $\mathrm{SU}(2)$ spin of a particle, this would manifest itself as a constant term in the `energy' of magnons (i.e. a chemical potential), and would lead to a term $\sim B\times P$ in the TBA equation for $Y_P$, cf. eqs. \eqref{eq:TBAgen}. Since $c\star s=c/2$ for constant $c$, such a term is in the null space of $(K+1)^{-1}_{PQ}$ (cf. eqn. \eqref{eq:Kplus1invdef}), and hence the simplified TBA equations would not distinguish between different values of this magnetic field. In short, the canonical TBA equations carry more information than the simplified TBA equations. We will not explicitly resolve this technical point here, but will briefly come back to it in section \ref{sec:YsystoTBA}.\footnote{Further discussion can be found in e.g. chapter four of \cite{vanTongeren:2013gva}.} The extra information required to reconstruct our magnetic field for example, lies in the large $u$ asymptotics of the Y functions, and upon specifying this information our simplified TBA equations are good to go.

The infinite sum in the main TBA equation can also be removed. Noting that similarly to $K^{fQ}$, $K^{Qf}$ satisfies
\begin{equation}
\label{eq:KMfidentity}
K^{Qf} - I_{QP} s \star K^{Pf}= s \delta_{Q1},
\end{equation}
we can rewrite the above simplified equations as
\begin{equation}
\log Y_Q - I_{QP} \log Y_P \star s = I_{QP} \log \left(1 + \frac{1}{Y_{P}}\right) \star s + \delta_{Q,1}\log \left(1 + \frac{1}{Y_f}\right) \star s.
\end{equation}
Integrating with $K^{Qf}$ and using eqn. \eqref{eq:KMfidentity} we get
\begin{equation}
\log Y_1 \star s = \log \left(1 + \frac{1}{Y_{Q}}\right) \star K^{Qf} - \log \left(1 + \frac{1}{Y_{1}}\right) \star s  + \log \left(1 + \frac{1}{Y_f}\right) \star s \star K^{1f},
\end{equation}
or in other words
\begin{equation}
\label{eq:infmagnonsumidentity}
\log \left(1 + \frac{1}{Y_{Q}}\right) \star K^{Qf} = \log \left(1 + Y_{1}\right) \star s  - \log \left(1 + \frac{1}{Y_f}\right) \star s \star K^{1f}.
\end{equation}
The main TBA equation \eqref{eq:cGNcanonicalTBA1} then becomes
\begin{equation}
\log Y_f = \frac{E}{T} - \log \left(1 + Y_{1}\right) \star s ,
\end{equation}
upon noting that magically enough the $Y_f$ contribution drops out completely thanks to $K^{ff}=s \star K^{1f}$.\footnote{To show this we can for example compute the integral in the second term by residues. The cancellation of the complicated scalar factor of the S matrix in the simplified TBA equations appears to be ubiquitous, an observation first made in \cite{Zamolodchikov:1991et}, as an intriguing manifestation of what must be crossing symmetry. Interestingly, at least in some cases we can reverse-engineer the scalar factor from this property \cite{Janik:2008hs}.} For uniformity we can define $Y_0 \equiv Y_f^{-1}$ and get
\begin{equation}
\label{eq:cGNsTBA}
\log Y_M = \log (1+Y_{M+1})(1+Y_{M-1}) \star s - \delta_{M,0}\left(\frac{E}{T}\right)
\end{equation}
with $Y_{M}\equiv 0$ for $M<0$.

\subsubsection*{Y system}

\begin{figure}%
\centering
\includegraphics[width=6cm]{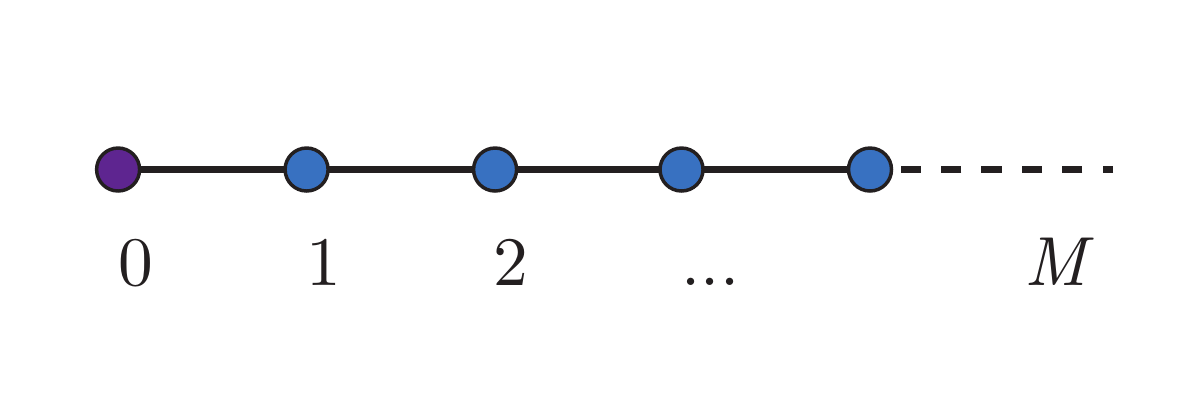}
\caption{The TBA structure for the chiral Gross-Neveu model in diagrammatic form. This graph illustrates the coupling between nearest neighbours in the simplified TBA equations \eqref{eq:cGNsTBA} or Y system \eqref{eq:cGNYsys}, where the different colour on the first node signifies the fact that it is `massive' corresponding to the $\delta_{M,0}$ term in the simplified equations (this is also frequently denoted by putting a $\times$ in the open circle).}
\label{fig:cGNYsys}
\end{figure}

\begin{figure}%
\centering
\includegraphics[width=6cm]{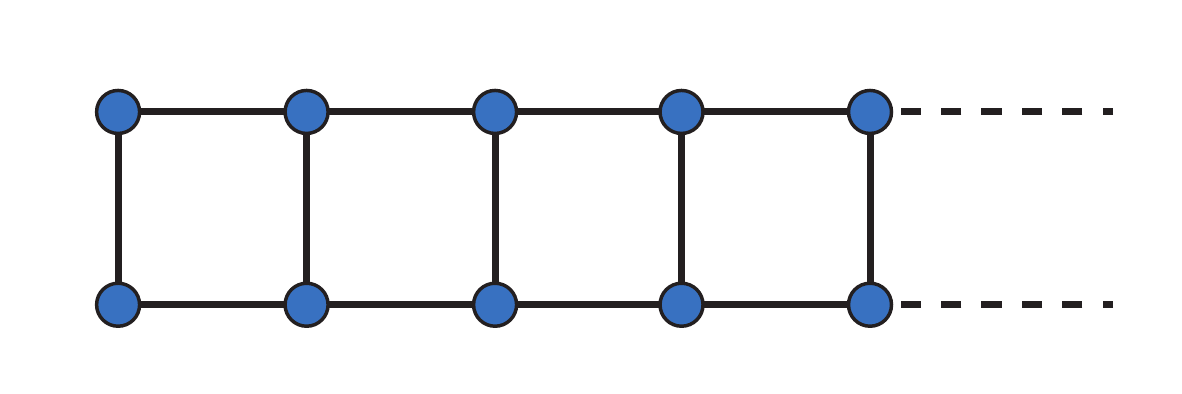}
\caption{The $\mathrm{SU}(3)$ Y system in diagrammatic form.}
\label{fig:cGNYsysSU3}
\end{figure}

To finish what we started, we can now apply $s^{-1}$ to these equations to get
\begin{equation}
\label{eq:cGNYsys}
Y^+_M \, Y^-_M  = (1+Y_{M+1})(1+Y_{M-1}),
\end{equation}
where the $\pm$ denote shifts in the argument by $\pm i$; $f^\pm(u) \equiv f(u\pm i)$. Note that the energy is in the null space of $s^{-1}$. These equations are known as the \emph{Y system} \cite{Zamolodchikov:1991et}. In general, the structure of simplified TBA equations and Y systems can be represented diagrammatically by graphs. For example, in this case eqs. \eqref{eq:cGNsTBA} and \eqref{eq:cGNYsys} can be represented by figure \ref{fig:cGNYsys}.
For more general models the Y system is defined on a certain two dimensional grid, for instance the $\mathrm{SU}(3)$ chiral Gross-Neveu model and $\mathrm{SU}(3)$ version of the Heisenberg spin chain would have a Y system corresponding to the diagram in figure \ref{fig:cGNYsysSU3}.
These diagrams have a group theoretical interpretation. We got extra Y functions for the XXX spin chain and chiral Gross-Neveu model due to the presence of bound states. These bound states of $Q$ particles carry total spin $Q/2$, which we can put into correspondence with the irreducible representations of $\mathrm{SU}(2)$. For higher rank symmetry algebras like $\mathrm{SU}(3)$, the story is similar: the Y functions correspond to inequivalent non-singlet irreducible representations. The irreducible representations of $\mathrm{SU}(3)$ can be represented by Young diagrams of maximal height three. All inequivalent non-singlet ones correspond to diagrams of height two, however, which match the entire diagram of figure \ref{fig:cGNYsysSU3} if we draw a square around every node.\footnote{There are also many integrable models with so-called quantum group symmetry. The representation theory in these cases is more involved, and for instance can result in a maximal spin. Correspondingly, in such cases TBA analysis results in a Y system with finitely many Y functions, see e.g. chapter 7 of \cite{vanTongeren:2013gva} for more details. An extensive review on Y systems and so-called T systems can be found in \cite{Kuniba:2010ir}.}

Let us emphasize again that in this process we lose information at each step along the way: both $(K+1)^{-1}$ and $s^{-1}$ have null-spaces. Therefore the simplified TBA equations are only equivalent to the canonical TBA equations provided we specify additional information on the Y functions such as their large $u$ asymptotics. An alternative but when applicable equivalent specification often encountered in the literature is to give the large $Q$ asymptotics of the $Y_Q$ functions.\footnote{Already for constant solutions of say the simplified TBA equations of the chiral Gross-Neveu model with $Y_0\rightarrow0$ there is large ambiguity: for constant Y functions the simplified TBA equations are equivalent to the Y system (of course without rapidities to shift), which is now nothing but a recursion relation fixing everything in terms of $Y_1$. As will come back below, only one value of this constant corresponds to a solution of the canonical equations with fixed chemical potentials.} The Y system requires even further specifications to really correspond to a particular model. For example the Y system for the XXX spin chain is given by dropping $Y_0$ from the chiral Gross-Neveu Y system altogether, but this is nothing but the chiral Gross-Neveu Y system again, just shifting the label $M$ by one unit.

\section{Integrability in finite volume}

\label{sec:intfinvol}

So far we have used integrability to get an exact description of the large volume limit of our theory, and used this to find a description of its thermodynamic properties in this limit. When the system size is finite however, the notion of an S-matrix -- let alone factorized scattering -- does not exist, making our integrability approach fundamentally inapplicable. Interestingly however, there is a way around this, allowing us to compute the finite size spectrum of an integrable field theory exactly. Parts of this section directly follow the corresponding discussion in chapter 2 of \cite{vanTongeren:2013gva}.

\subsection{The ground state energy in finite volume}

\begin{figure}%
\centering
\includegraphics[width=0.7\textwidth]{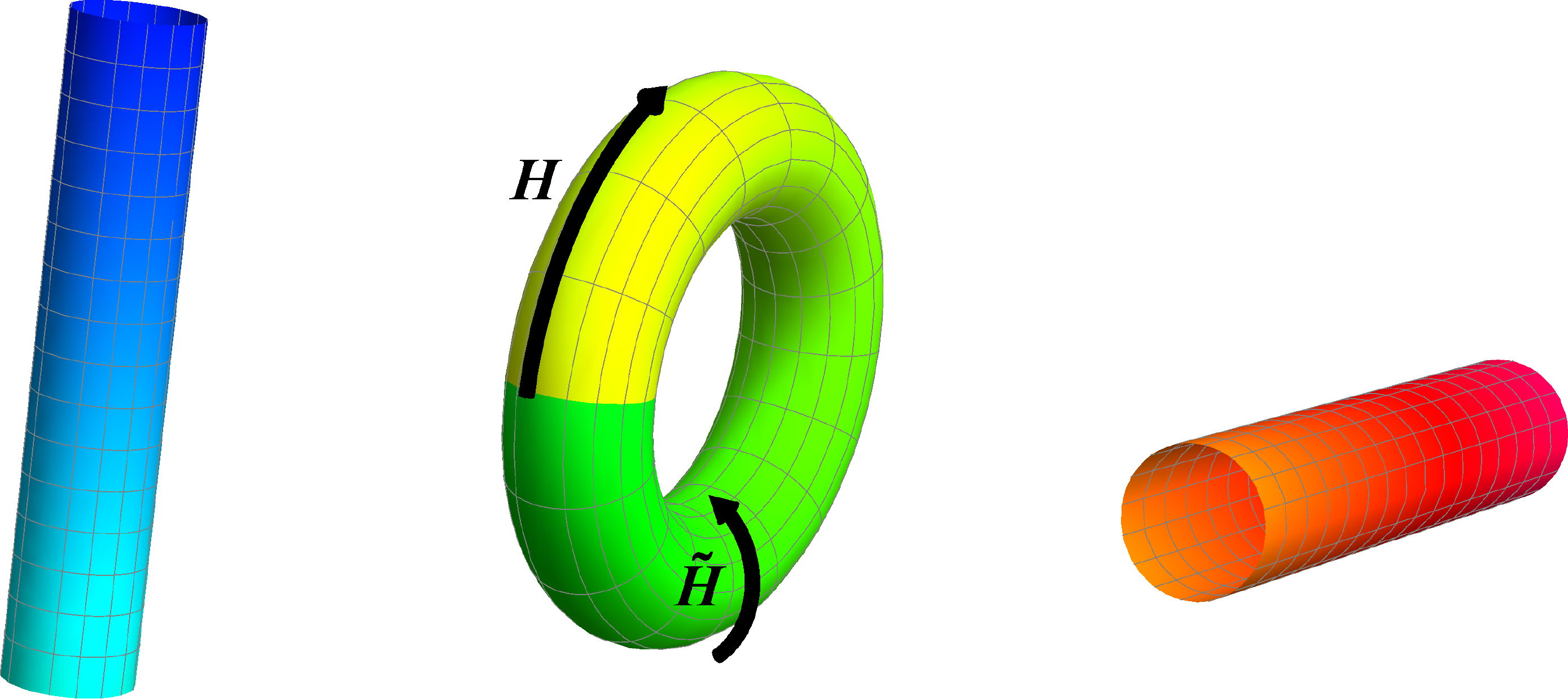}
\caption{The mirror trick. The partition function for a theory on a finite circle at finite temperature lives on a torus (middle). In the zero temperature limit this torus degenerates and gives the partition function on a circle at zero temperature (left), dominated by the ground state energy. Interchanging space and time we obtain a mirrored view of this degeneration as the partition function of the mirror theory at finite temperature but on a decompactified circle, determined by the infinite volume mirror free energy (or Witten's index).}
\label{fig:mirrortrick}
\end{figure}

Let us not be too ambitious and begin by attempting to compute the ground state energy of our theory in finite volume. This is possible thanks to a clever idea by Zamolodchikov \cite{Zamolodchikov:1989cf}. To describe this idea let us recall that the ground state energy is the leading low temperature contribution to the (Euclidean) partition function
\begin{equation}
\label{eq:ZlimittoGSE}
Z(\beta,L) = \sum_n e^{-\beta E_n} \sim e^{-\beta E_0} , \, \hspace{20pt} \mbox{as} \, \, \beta \equiv \frac{1}{T} \rightarrow \infty.
\end{equation}
We can compute this partition function with our original quantum field theory by Wick rotating $\tau \rightarrow \tilde{\sigma} =i \tau $ and considering a path integral over fields periodic in $\tilde{\sigma}$ with period $\beta$. Geometrically we are putting the theory on a torus which in the zero temperature limit degenerates to the cylinder we began with. Analytically continuing $\tilde{\sigma}$ back to $\tau$ gives back our original Lorentzian theory. We could, however, also analytically continue $\sigma \rightarrow \tilde{\tau} = -i \sigma$. This gives us a Lorentzian theory where the role of space and time have been interchanged with respect to the original model -- it gives us its \emph{mirror model}.\footnote{A double Wick rotation leaves a relativistic field theory invariant, and hence we do not really need to carefully make this distinction here. Still, we will occasionally do so for pedagogical purposes. After gauge fixing the integrable models encountered in the context of the AdS/CFT correspondence are not Lorentz invariant for instance, meaning the double Wick rotation produces a different model. The term mirror model and mirror transformation were introduced in this context in \cite{Arutyunov:2007tc}. Interestingly, the $\ads$ mirror model -- the model on which the AdS$_5$/CFT$_4$ TBA is based -- can be interpreted as a string itself \cite{Arutyunov:2014cra,Arutyunov:2014jfa}. The spectrum of this string is thereby related to the thermodynamics of the $\ads$ string, and vice versa \cite{Arutynov:2014ota}.} Putting it geometrically, we could consider Hamiltonian evolution along either of the two cycles of the torus. Note that at the level of the Hamiltonian and the momentum the mirror transformation corresponds to
\begin{equation}
H \rightarrow i \tilde{p}, \hspace{20pt} p \rightarrow - i \tilde{H},
\end{equation}
where mirror quantities are denoted with a tilde. To emphasize its role as the mirror volume, let us from now on denote the inverse temperature $\beta$ by $R$. In principle we can compute the Euclidean partition function both through our original model at size $L$ and temperature $1/R$ and through the mirror model at size $R$ and temperature $1/L$. These ideas are illustrated in figure \ref{fig:mirrortrick}.

To find the ground state energy of our model then, we could equivalently compute the infinite volume partition function of our mirror model at finite temperature, i.e. its (generalized) free energy $\tilde{F}$ since
\begin{equation}
Z = e^{-L \tilde{F}}.
\end{equation}
In fact, cf. eqn. \eqref{eq:ZlimittoGSE}, the ground state energy is related to the free energy density of the mirror model as
\begin{equation}
E_0 = \frac{L}{R} \tilde{F} = L \tilde{f}.
\end{equation}
The key point of this trick is that we are working with the mirror model in the infinite volume limit where we can use factorized scattering and the asymptotic Bethe ansatz of the previous section, since any exponential corrections to them can be safely neglected.\footnote{Note again that the mirror of a relativistic model is equal to the original (up to the specific boundary conditions required to compute the same partition function), and therefore the mirror model is immediately integrable as well. In general the conservation laws responsible for factorized scattering are preserved by Wick rotations, so that the mirror theory has many conserved quantities and mirror scattering should factorize. Moreover we can obtain the S-matrix from four point correlations functions via the LSZ reduction formula, and correlation functions can be computed by Wick rotations.} The price we have to pay is dealing with a finite temperature. Fortunately we just learned how to do precisely this, and we can compute our ground state energy from the thermodynamic Bethe ansatz applied to the double Wick rotated (mirror) model.

We should be a little careful about the boundary conditions in our model however. Where fermions are concerned the Euclidean partition function is only the proper statistical mechanical partition function used above, provided the fermions are anti-periodic in imaginary time. Turning things around, if the fermions are periodic on the circle then from the mirror point of view they will be periodic in imaginary time, so that our goal in the mirror theory is not to compute the standard statistical mechanical partition function but rather what is known as Witten's index
\begin{equation}
Z_W = \mbox{Tr} \left( (-1)^F e^{-L \tilde{H}}\right),
\end{equation}
where $F$ is the fermion number operator. This means we are adding $i \pi F/L$ to the Hamiltonian -- a constant imaginary chemical potential for fermions.\footnote{\label{footnote:QPBC}Continuing along these lines, if we were to consider quasi-periodic boundary conditions instead of (anti-)periodic boundary conditions a more general operator enters in the trace, which leads to more general chemical potentials. For details see e.g. chapter two and four of \cite{vanTongeren:2013gva}.}

We should also note that the mirror transformation actually has a nice meaning on the rapidity plane, provided we adapt it slightly. From our discussion above, we see that the energy and momentum of a particle should transform as
\begin{equation}
E \rightarrow i \tilde{p}, \hspace{20pt} p \rightarrow - i \tilde{E},
\end{equation}
which leaves its relativistic dispersion relation $E^2 - p^2 = m^2$ invariant. This means we can parametrize $\tilde{E}$ and $\tilde{p}$ exactly as before ($E(u)=\cosh \tfrac{\pi u}{2}$ and $p(u)=\sinh \tfrac{\pi u}{2}$), but let us say now in terms of a mirror rapidity $\tilde{u}$. We can then wonder what the relation between $u$ and $\tilde{u}$ should be. By definition we want
\begin{equation}
E(u)= i p(\tilde{u}), \hspace{20pt} p(u)= -i E(\tilde{u}).
\end{equation}
Now we recall that sines and cosines are related by shifts of $\pi/2$, which in the hyperbolic case tells us that
\begin{equation}
E(u-i) = i \sinh\tfrac{\pi u}{2}  = -i p(-u), \hspace{20pt} p(u-i) = -i \cosh\tfrac{\pi u}{2}  = -i E(\pm u).
\end{equation}
Hence we see that if we identify $-\tilde{u}=u-i$, we get what we want. In the literature you will however typically encounter the transformation $u \rightarrow u+i$ ($\theta \rightarrow \theta + i\tfrac{\pi}{2}$ in the standard relativistic rapidity parametrization ) which is quite convenient and we will use from here on out.\footnote{While widely used, the name mirror transformation is appropriate for the case we started with, as you can readily convince yourself of by drawing a picture in the complex $(\sigma,\tau)$ plane. What does the second transformation do?} Here the rapidity on the right hand side actually implicitly refers to the mirror rapidity $\tilde{u}$, matching our story so that
\begin{equation}
u \rightarrow \tilde{u}+i, \mbox{ i.e. } \tilde{u} = u-i.
\end{equation}
This means that in addition to what we are doing here, people frequently do a parity transformation in between. For parity invariant theories this does absolutely nothing, and even if a theory is not parity invariant, we could simply proceed this way and compute things in the parity flipped theory, reverting back only at the final stage.

Applying the above discussion to the chiral Gross-Neveu model, we see that we can compute its ground state energy on a circle of circumference $L$ by taking our derivation of the free energy above, replacing the length $L$ by the mirror length $R$, replacing the temperature $T$ by the inverse length $1/L$, and adding a constant term $i \pi /L$ to the dispersion relation for the fermions. The ground state energy is then given by
\begin{equation}
\label{eq:TBAgroundstateenergy}
E_0 = - \int_{-\infty}^\infty du \frac{1}{2\pi} \frac{dp}{du} \log\left(1+Y_0\right),
\end{equation}
where $Y_0$ satisfies the (mirror) TBA equations
\begin{equation}
\label{eq:cGNsTBAGS}
\log Y_M = \log (1+Y_{M+1})(1+Y_{M-1}) \star s - \delta_{M,0}(L E +i \pi),
\end{equation}
together with the $Y_{M>0}$. Note the added $i \pi$ in line with the periodicity of the fermions in imaginary mirror time.

\subsection{Tricks with analytic continuation}

At this point we have actually done something quite impressive: we have found a system of equations we can solve (admittedly numerically) to find the exact finite volume ground state energy of a two dimensional field theory. It would be great if we could extend this approach to the entire spectrum. If we look back at our arguments however, we are immediately faced with a big conceptual problem; the mirror trick and infinite volume limit work nicely precisely for the ground state and the ground state only! Still it is hard to believe that a set of complicated TBA equations knows about the ground state only, especially since they are derived from the mirror Bethe-Yang equations which are just an analytic continuation away from describing the \emph{complete} large volume spectrum. In this section we will take an approach often taken in physics; we will (try to) analytically continue from one part of a problem to another, in this case from the ground state energy to excited state energies. The idea that excited states can be obtained by analytic continuation is an old one, discussed in e.g. \cite{Bender:1969si} in the case of the quantum anharmonic oscillator.

\subsubsection{A simple example}

\begin{figure}%
\centering
\subfigure[]{\includegraphics[width=7cm]{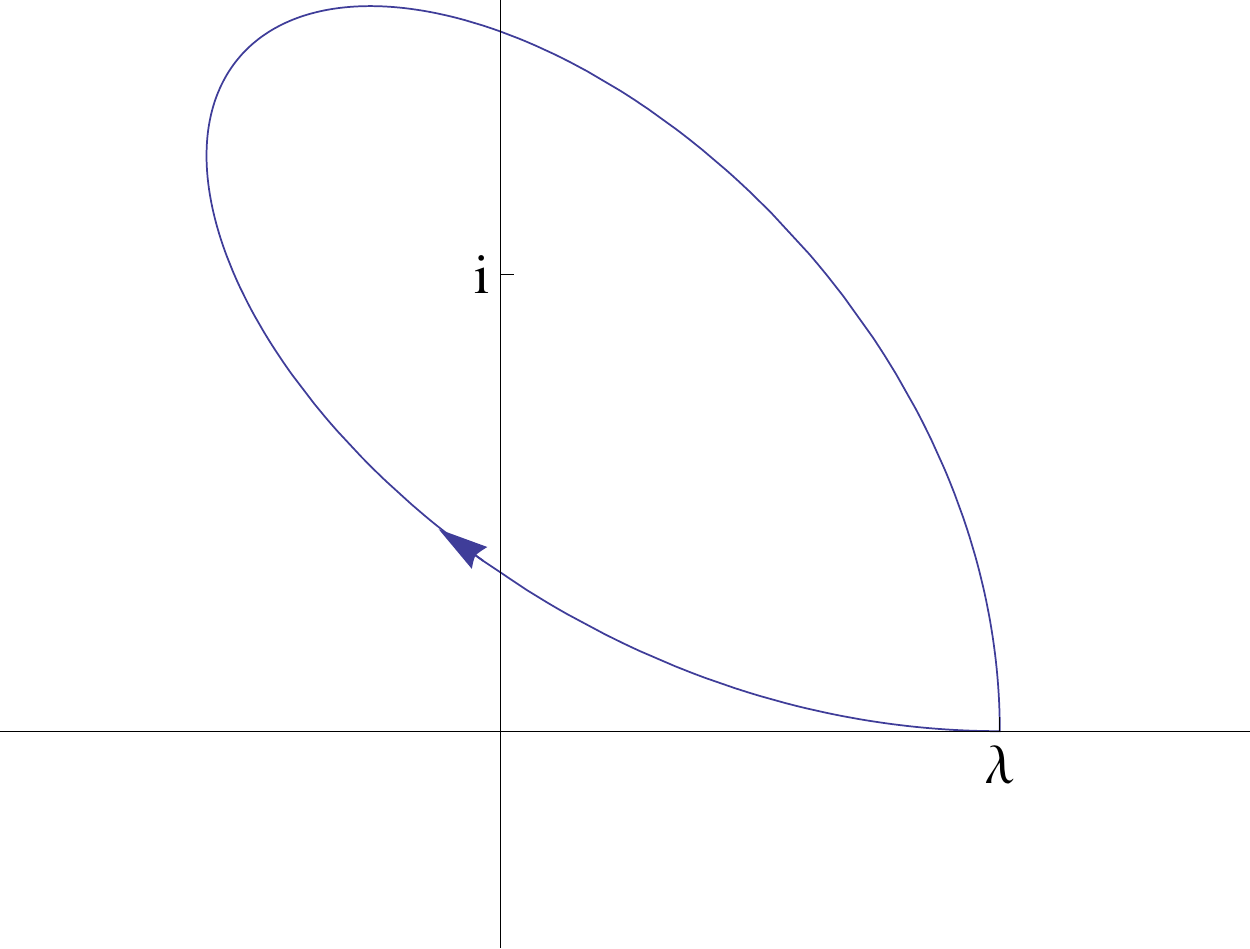} \label{fig:aroundthebranchpoint}} \quad
\subfigure[]{\includegraphics[width=7cm]{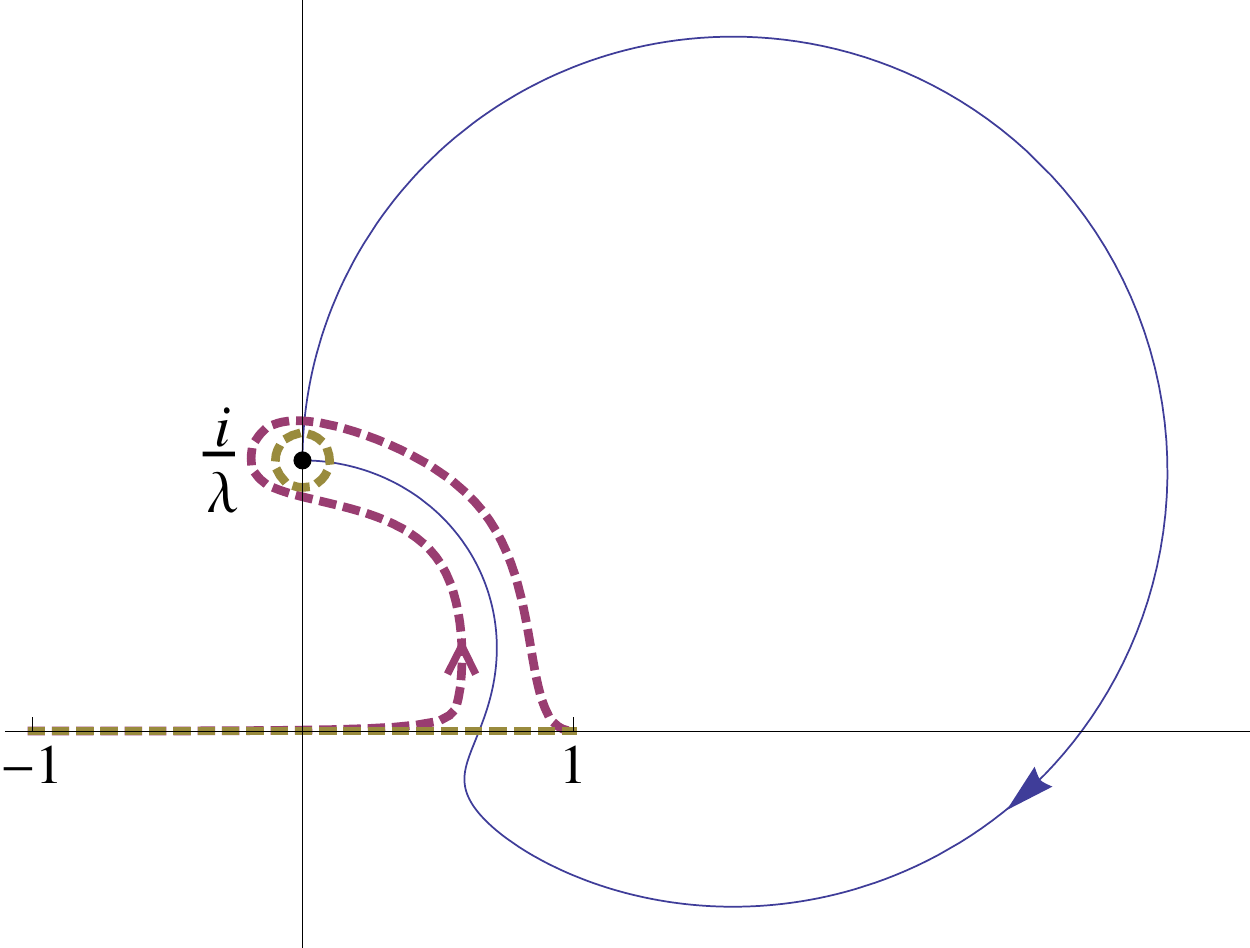} \label{fig:throughthecontour}}
\caption{Analytic continuation. The left figure shows the analytic continuation of $\lambda$ (blue) around the branch point at $i$, corresponding to flipping the sign of $-\sqrt{1+\lambda^2}$ upon returning to the real line. The right figure shows the corresponding movement of the pole at $i/\lambda$ (blue) which drags the integration contour (red, dashed) in eqn. \eqref{eq:simpleenergyasintegral} along with itself for continuity. Upon taking the integration contour back to the real line we retain a residual contribution (yellow, dashed).}
\label{fig:analyticcont}
\end{figure}

Before moving on, we would like to motivate these ideas and illustrate them on a simple quantum mechanical problem\footnote{This nice example can be found in slides of a talk by P. Dorey at IGST08 \cite{doreyIGST08}.}
\begin{equation}
H \psi = E \psi , \,\,\,\,\, \mbox{with} \, \, \, \,\, H = \left(\begin{array}{cc} 1 & 0 \\ 0 & -1 \end{array}\right) + \lambda \left(\begin{array}{cc}0 & 1 \\ 1 & 0 \end{array}\right).
\end{equation}
After considerable effort we realize that the spectrum in this model is given by
\begin{equation}
E(\lambda) = \pm\sqrt{1+\lambda^2},
\end{equation}
and hence the ground state energy is $-\sqrt{1+\lambda^2}$. Allowing ourselves to analytically continue in the coupling constant we realize that the equation for the ground state energy has branch points at $\lambda = \pm i$. As a consequence, analytically continuing around either of these branch points and coming back to the real line we do not quite get back the ground state energy, but rather the energy of the excited state. This is illustrated in figure \ref{fig:aroundthebranchpoint}.
The message we can take away from this \cite{doreyIGST08} is that by analytically continuing a parameter around a ``closed contour'' -- meaning we come back to the ``same'' value though not necessarily on the same sheet -- we end up back at the same problem although our eigenvalue may have changed. As we are still dealing with the same problem, if the eigenvalue has changed under analytic continuation it must have become one of the other eigenvalues. Note that this does not imply all eigenvalues can be found this way -- the spectrum may split into distinct sectors closed under analytic continuation.

Let us now forget about the description of this problem in terms of linear algebra, and suppose for the sake of the argument that in solving our spectral problem we had obtained
\begin{equation}
\label{eq:simpleenergyasintegral}
E(\lambda) = -\int_{-1}^{1} dz \, \frac{1}{2\pi i} f(z) g(z)-1,
\end{equation}
where
\begin{equation}
f(z) = \frac{1}{z - i/\lambda},\,\,\,\, \mbox{and} \,\,\,\, g(z) = 2 \lambda \sqrt{1-z^2}.
\end{equation}
We can determine that this integral has branch points at $\lambda = \pm i$ without knowing anything about $f(z)$ other than that it is meromorphic with a single pole at $i/\lambda$. Conceptually we consider $g(z)$ to be some nice known function, while $f$ is not explicitly known. Analytically continuing the integral in $\lambda$ we get a function that is well defined everywhere except for the half-lines $i \lambda>1$ and $i \lambda<-1$ where the pole moves into the integration domain. Continuing around the point $\lambda = i$ as in figure \ref{fig:aroundthebranchpoint}, nothing happens when we first cross the line $\mbox{Re}(\lambda)=0$ but when we cross the second time, the pole moves through the integration contour on the real line and drags the contour along, as illustrated in figure \ref{fig:throughthecontour}. We can rewrite the resulting contour integral in terms of the original one by picking up the residue, giving
\begin{equation}
E^c(\lambda) = -\int_{-1}^{1} dz \, \frac{1}{2\pi i} f(z) g(z) + g(i/\lambda)-1.
\end{equation}
Since $E^c(\lambda) - E(\lambda) = g(i/\lambda) \neq 0$ there must be a branch point inside the contour. In this integral picture we do not need to know the precise analytic expression of $E$ or $f$ to determine the expression for the excited state energy. All we need to know is the pole structure of $f$ relative to the integration contour.

\subsubsection{Analytic continuation of TBA equations}

Inspired by this example, we can try to analytically continue our expression for the ground state energy, eqn. \eqref{eq:TBAgroundstateenergy}, in some appropriate variable and see whether we encounter any changes in the description. We could try continuing in the mass variable of the chiral Gross-Neveu model for example. This approach to excited states in the TBA was proposed and successfully applied to what is known as the scaling Lee-Yang model in \cite{Dorey:1996re}. The authors there observed that in the process of analytic continuation the Y functions solving the TBA equations undergo nontrivial monodromies. They moreover noted that changes in the form of the TBA equations are possible if singular points of $1+1/Y$ move in the complex plane during the analytic continuation. These changes are analogous to the changes in the energy formula of our example above. Here the integral is a typical term on the right hand side of the TBA equations
\begin{equation}
y(u) \equiv \log\left(1 + \frac{1}{Y}\right)\star K(u),
\end{equation}
where we recall that $\star$ denotes (right) convolution on the real line. If there is a singular point
\begin{equation}
\label{eq:singularpoint}
Y(u^*) = -1,
\end{equation}
and its location $u^*$ crosses the real line during the analytic continuation, we can pick up the residue just as in our simple example to get
\begin{equation}
y^c(u) = \log\left(1 + \frac{1}{Y}\right)\star K(u) \pm \log S(u^*,u),
\end{equation}
where we recall that $K(v,u) = \frac{1}{2\pi i} \frac{d}{dv} \log S(v,u)$ and the sign is positive for singular points that cross the contour from below and negative for those that cross it from above. If Y vanishes at a particular point, this leads to the same considerations, just resulting in an opposite sign.\footnote{The singular points of different Y functions in the complex plane are typically related. A driving term arises from a special point $u^*$ for a Y function on the right hand side of a TBA equation. Since this term typically has poles at $u^*\pm i$, however, this shifted point corresponds to a zero or pole for the Y function on the left hand side. Analyzing the Y system (discussed just below) we arrive at the same conclusion.} If we wanted to do this at the level of the simplified equations, all we need is the S-matrix associated to $s$:
\begin{equation}
\label{eq:canonicalSmatrix}
S(u) = - \tanh \frac{\pi}{4}(u-i).
\end{equation}
The energy itself is also determined by an integral equation in the TBA approach, meaning it can change explicitly as well as implicitly through the solution of a changed set of TBA equations.

The upshot of this is that we obtain excited state TBA equations that differ from those of the ground state by the addition of $\log S$ terms, which we will call \emph{driving terms}. It should not matter whether we consider this procedure at the level of the canonical equations or at the level of the simplified equations, and indeed the results agree because of the S-matrix analogue of identities like eqn. \eqref{eq:KQMidentity}.

\subsection{Excited states and the Y system}
\label{sec:YsystoTBA}

The case of the Y system is a bit more peculiar, since the distinguishing features of an excited state completely disappear. This is because the S-matrix \eqref{eq:canonicalSmatrix} vanishes under application of $s^{-1}$. From this we see that whatever excited state TBA equations we obtain by the above reasoning, the Y system equations are the same as those of the ground state: the Y system is \emph{universal}.\footnote{Exemptions to this rule can arise under very specific circumstances, see e.g. \cite{Arutyunov:2012ai} and chapter seven of \cite{vanTongeren:2013gva}.} The important distinction is that as we just said the Y functions for excited states have singular points. If there are no further singularities like branch cuts (which we would expect to be universal features of a model rather than state dependent), specifying the number of simple poles and zeroes of all Y functions in the strip between $i$ and $-i$ is almost enough to `integrate' the Y system back to the simplified TBA equations. First, however, we need to address the fact that different physical models can have the same Y system. This also brings us back to the discussion of information loss in the simplifying steps of section \ref{sec:TBAtoY}.

\subsubsection*{Asymptotics of Y functions}

For concreteness, let us consider the simplified TBA equations \eqref{eq:cGNsTBA} for the chiral Gross-Neveu model. The distinguishing feature of these equations with respect to say the XXX ones is the energy contribution to $Y_0$. This term leads to $\log Y_0 \sim - e^{\pi/2 |u|}/T$ at large $|u|$, meaning $Y_0$ goes to zero quite rapidly. If we take these asymptotics as given and assume $Y_0$ is analytic in the strip between $i$ and $-i$, for the time being interpreting $Y_1$ as some given external function, we can `integrate' the Y system equation $Y_0^+ Y_0^- = (1+Y_1)$ to the associated simplified TBA equation. Namely
\begin{equation}
Y_0 = e^{-E/T} e^{\log(1+Y_1)\star s}
\end{equation}
satisfies the Y system (note again that $e^{-E/T}$ drops out of this), has the right asymptotics, and is analytic, which by Liouville's theorem means it is unique (the difference with any other function with these properties is zero).

To get the simplified TBA equations for the remaining Y functions, which have no energy terms, it turns out we should demand constant asymptotics $Y_N \rightarrow \hat{Y}_N$. These constants are all recursively determined by one of them, e.g. $\hat{Y}_1$, by the constant limit of the Y system (where $\hat{Y}_0=0$ in line with its asymptotics), i.e.
\begin{equation}
\begin{aligned}
\hat{Y}_2 & = \hat{Y}_1^2-1,\\
\hat{Y}_{N+1} & = \frac{\hat{Y}_N^2}{1+\hat{Y}_{N-1}} -1, \,\,\,\,\, N>1.
\end{aligned}
\end{equation}
A simple solution to this set of equations is $Y_M = M(M+2)$, essentially due to the identity $M^2 = (M+1)(M-1)+1$. We can generalize this solution to
\begin{equation}
\hat{Y}_M = [M]_q [M+2]_q,
\end{equation}
where we introduced the so-called $q$ numbers
\begin{equation}
[M]_q  = \frac{q^M - q^{-M}}{q-q^{-1}},
\end{equation}
which retain the property $[M]_q^2 = [M+1]_q[M-1]_q+1$ for any $q \in \mathbb{C}$. In the limit $q\rightarrow1$, $[N]_q \rightarrow N$ again. Since everything is recursively fixed by $\hat{Y}_1 = [3]_q$ and by picking $q$ appropriately we can make $[3]_q$ any complex constant, this is the general constant solution of our Y system. Given a value of $\hat{Y}_1$ and hence all $\hat{Y}_M$, the expression for the associated full Y functions as the right hand sides of their TBA equations follows uniquely from analyticity and the Y system, as it did for $Y_0$.

To fix the constant asymptotic of $Y_1$ we can feed our constant ``solution'' in to the canonical TBA equations, where now integration with the kernels amounts to multiplication by their normalizations. Then performing the infinite sums in the canonical TBA equations we get a set of equations that admits only one value for $\hat{Y}_1$. In our chiral Gross-Neveu case this fixes the asymptotes of $Y_{M>0}$ to be $M(M+2)$. If we had included a nontrivial chemical potential $\mu$ for the spin down fermions in a thermodynamic picture, or double Wick rotated quasi-periodic boundary conditions as in footnote \ref{footnote:QPBC}, we would instead be required to take a different constant $q$ number solution with $\log q \sim \mu$, showing the physical interpretation of these constant asymptotics.\footnote{More details can be found in e.g. chapters 2 and 4, and appendix A.4 of \cite{vanTongeren:2013gva}. In particular, evaluating the infinite sums actually requires an $i \epsilon$ prescription in case of nonzero chemical potential.} This link between chemical potentials and asymptotics actually allows us to move between canonical TBA equations and simplified equations plus (constant) asymptotics.

As mentioned earlier, the XXX spin chain has the same Y system, but different (simplified) TBA equations. These simplified TBA equations would follow along the same lines, but with different asymptotics. Similarly, the $i\pi$ contribution in eqs. \eqref{eq:cGNsTBAGS} affects the asymptotics relative to eqs. \eqref{eq:cGNsTBA}.

\subsubsection*{Poles and zeroes of Y functions}

Now that we have seen how to get basic simplified TBA equations from a Y system, let us try to add driving terms. To do so, we need to know the simple poles and zeroes of the Y functions. Provided we are given this data, we can explicitly factor out poles and zeroes of $Y$ via products of $t(u) = \tanh\tfrac{\pi}{4} u$ and $1/t$. In other words for a Y function with poles at $\xi_i$ and zeroes at $\chi_j$ we define
\begin{equation}
\check{Y}(u) = \frac{\prod_{j} t(u-\chi_j)}{\prod_{i} t(u-\xi_i)}Y(u),
\end{equation}
which is analytic. We now start from the schematic Y system $Y^+ Y^- = \mathcal{R}$, which implies also $\check{Y}^+ \check{Y}^- = \mathcal{R}$ because $t^+ t^- = 1$. Morever, since $\check{Y}$ is analytic and has the same asymptote as $Y$ because $t(u)$ asymptotes to one, we are essentially in the situation we had above (the relation of $t(u)$ to $S(u)$ of eqn. \eqref{eq:canonicalSmatrix} is not accidental). By our previous analysis we get
\begin{equation}
\check{Y} = e^{\log \mathcal{R} \star s},
\end{equation}
so that
\begin{equation}
Y =  \frac{\prod_{i} t(u-\xi_i)}{\prod_{j} t(u-\chi_j)} e^{\log \mathcal{R} \star s},
\end{equation}
where we should include $e^{-E/T}$ as before if necessary. This is precisely of the form of a simplified excited state TBA equation. To reiterate, this formula by definition gives the Y system upon applying $s^{-1}$, and has the right poles, zeroes, and asymptotics, making it our unique desired answer. For more complicated TBA equations with branch cuts we would need to know the discontinuities of the Y functions across the cuts, in addition to poles, zeroes and asymptotics, but morally we would do the same thing.

In short, by supplying analyticity data in the form of poles, zeroes, and asymptotics, we can derive a set of integral equations of simplified TBA form, with precisely the expected type of energy and driving terms. Some form of integral equations is of course useful, as they can typically be iteratively solved, perhaps by starting from a seed solution in some part of parameter space (an asymptotic solution), which should in particular include appropriate starting values for the zeroes and poles. The notion that analytic properties might ``label'' excited states also appears in e.g. the discussion of the ``Bethe ansatz'' for the harmonic oscillator in the article by F. Levkovich-Maslyuk \cite{Levkovich-Maslyuk:2016kfv}.

The Y system and its universality are closely related to other approaches of obtaining equations that describe excited state energies. In some cases it is possible to construct a functional analogue of the Y system directly, as discussed in the article by S. Negro \cite{Negro:2016yuu}. If we can then get satisfactory insight into the analytic structure of the corresponding objects, we can `integrate' these functional relations in the above spirit to obtain integral equations describing the energy of excited states \cite{Klumper:1992vt,Klumper:1993,Bazhanov:1996aq,Fioravanti:1996rz}. Depending on how these functional equations are `integrated' we can obtain equations of TBA form but also various other forms that can be more computationally efficient. The latter equations generically go under the name of ``non-linear integral equations'' \cite{Klumper:1992vt}, but depending on the context are also called ``Kl\"umper-Pearce'' \cite{Klumper:1991-1,Klumper:1991-2,Klumper:1992vt,Klumper:1993vq} or ``Destri-de Vega'' \cite{Destri:1992qk,Destri:1994bv} equations. While not obvious from their form, when different types of equations are possible they should of course be equivalent \cite{Juttner:1997tc,Kuniba:1998}.

\subsection{L\"uscher formulae}
\label{sec:luscherandexcitedstatesgeneral}

In general, we may wish to use an amalgamation of the above ideas to find excited state TBA equations, in the form of something which we will refer to as the contour deformation trick. The basic idea goes as follows. We will find a candidate solution of the Y system for an excited state with some limited regime of applicability. We then assume that the form of the TBA equations for an excited state is uniform and does not change outside of the regime of applicability of our candidate solution. Next, drawing lessons from the analytic continuation story above we expect that the only changes in the equation should be the addition of possible driving terms. Furthermore, although our limited solution only gives us a static picture, we expect that we can qualitatively view these terms as if coming from singular points that crossed the integration contour. Since in this picture such singular points would have dragged the contour along with them, we expect that an excited state TBA equation should be of the same form as the ground state, except with modified integration contours. Analyzing the analytic structure of the candidate solution will allow us to consistently define these contours in such a way that the TBA equations are satisfied, and by taking the integration contours back to the real line we can explicitly pick up the corresponding driving terms. Coming back to our simple example, it would be as if internal consistency of the problem (perhaps in the form of some other equation) told us that the natural integration contour for the excited state was not the interval $(-1,1)$, but a contour that starts at one and finishes at minus one while enclosing $i/\lambda$ between itself and the real line. Such a contour is of course equivalent to the red contour in figure \ref{fig:throughthecontour} obtained by direct analytic continuation.

\begin{figure}[h]
\begin{center}
\includegraphics{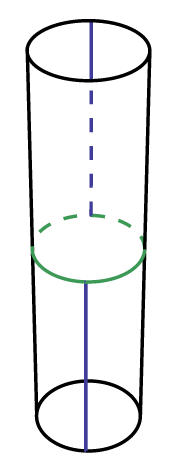}\hspace{50pt}\includegraphics{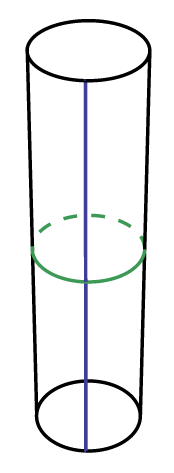}
\caption{The L\"uscher $\mu$- and F-term. On the left we have the decay of a physical particle (blue) into a pair of virtual particles (green) which fuse to a physical particle on the other side of the cylinder, while the right shows the scattering of a virtual particle with the physical particle as it loops around the cylinder.}
\label{fig:luscherprocesses}
\end{center}
\end{figure}

Through a bit of physical reasoning we can obtain a candidate solution of the TBA equations that should describe an excited state. If we take our theory at face value as a field theory on a cylinder, it is natural to expect the energy of states to get corrections from virtual particles travelling around the circle, a phenomenon investigated in particular by L\"uscher \cite{Luscher:1985dn}. Concretely, L\"uscher showed how polarization effects lead to mass corrections for a standing particle in massive quantum field theory in a periodic box, computing their effect to leading order in $e^{-m L}$ where $m$ is the mass of the particle and $L$ the size of the periodic box \cite{Luscher:1985dn}. These (leading order) corrections come in two types illustrated in figure \ref{fig:luscherprocesses}. The first of these is the so-called $\mu$ term corresponding to the particle decaying into a pair of virtual particle which move around the circle (in two dimensions) and recombine, while the second is the F-term which corresponds to a virtual particle loop around the circle which involves scattering with the physical particle.

Generalizing these ideas to moving particles and interacting multi-particle states based on the original diagrammatic methods of \cite{Luscher:1985dn} seems daunting. In the context of simple relativistic integrable models, however, L\"uscher's formulae readily follow by explicitly expanding the TBA equations in the large volume limit. By carefully generalizing the expansions in such models to interacting multi-particle states we can try to obtain a type of generalized L\"uscher's formulae. To leading order this energy correction takes the form \cite{Bajnok:2008bm}
\begin{equation}
\label{eq:deltaEluscherleading}
\Delta E = - \sum_Q \int_{-\infty}^{\infty} \frac{d\tilde{p}}{2\pi} e^{-\tH L }\lambda_{Q,1}(\tilde{p}|\{p_j\}),
\end{equation}
where we have indicated double Wick rotated (mirror) quantities by a tilde to show their origin, though the distinction will not matter for us here. This is the multi-particle generalization of the contribution corresponding to the F term on the right side of figure \ref{fig:luscherprocesses}. In many integrable models the $\mu$ term does not appear to show up at leading order for most states. In this formula, $\lambda_{Q,1}(\tilde{p}|\{p_j\})$ denotes the eigenvalue of the transfer matrix for the state of the integrable model under consideration, with its auxiliary space taken to be the mirror (double Wick rotated) representation for a mirror particle of type $Q$. In other words, the energy shift is given by scatter any possible virtual particle\footnote{At least pictorially it is clear that a virtual particle is like a regularly propagating mirror particle.} with the physical excitations of our large volume state, and summing over all of them, weighed by $e^{-\tH L }$.

Now we argued above that the excited state TBA equations should differ from the ground state ones by a set of driving terms, but should otherwise be of the exact same form. Considering the energy formula \eqref{eq:TBAgroundstateenergy} in this light, we realize that at large mass or large volume the $Y_f$ function should be small due to the $-L \tH  = - m L \cosh{\pi u/2}$ term in their canonical TBA equations. Expanding the energy formula for small $Y_0$ and comparing this to the leading weak coupling correction \eqref{eq:deltaEluscherleading}, where for the chiral Gross-Neveu model there is no sum over $Q$ since there are no physical bound states, for an excited state described by a set of rapidities $\{u_i\}$ it is natural to identify
\begin{equation}
Y_0^o(\tilde{u}) = e^{-E(\tilde{u}) L} \prod_{i=1}^\NF S^{ff}(\tilde{u},u_i) \lambda(\tilde{u}|\{u_i\}),
\end{equation}
where the tilde is a label to indicate that the associated quantities are to be evaluated in the mirror theory, and $\lambda$ refers to the XXX spin chain transfer matrix eigenvalue without the scalar factor $S^{ff}$, which we hence have to reinstate to describe our chiral Gross-Neveu model. The superscript $o$ indicates that this is an asymptotic solution that only applies to leading order in $e^{-E L}$. One immediate promising feature of this formula is that if we analytically continue this function from the mirror theory to the physical theory we are interested in, this precisely looks like the right hand side of the Bethe equations, and we get that asymptotically
\begin{equation}
Y_0^o(u^*_i) = -1,
\end{equation}
the $*$ denoting that we have analytically continued. This precisely corresponds the kind of singular point we encountered in our general discussion around eqn. \eqref{eq:singularpoint}! In fact, assigning appropriate driving terms to these singular points precisely results in an energy formula of the form
\begin{equation}
\label{eq:TBAgeneralstateenergyformula}
E = \sum_{i=1}^{\NF} E(p_i) -\int du \, \frac{1}{2\pi}\frac{dp}{du}\log\left(1+Y_0 \right)\,,
\end{equation}
where $E(p_i)$ is the asymptotic energy of the $i$th particle (recall that $p$ evaluated on an analytically continued rapidity is just $-i E$). Of course there can be further modifications to this energy formula depending on possible further singular points of $Y_0$, see for instance \cite{Arutyunov:2011mk} for a situation with rather involved analytic properties. Actually, the auxiliary equations could change as well, leading us to wonder what asymptotic solution we should consider there. Not going into technical details, we hope the following sounds reasonable. The auxiliary $Y_Q$ functions are physically associated to the Bethe-string solutions of the XXX spin chain (with inhomogeneities), which as we discussed are bound states, and their S-matrices can be found by fusion. We could construct transfer matrices based on each of the bound state S-matrices labeled by the string length $P$, and find their eigenvalues. By construction these objects will satisfy a relation similar to, but slightly more complicated than the one for the diagonalized scattering amplitudes of eqn. \eqref{eq:Sdiscretelaplace}, and by using these relations one can consistently express the $Y_{P>0}$ in terms of these bound state transfer matrix eigenvalues. This then gives us a full asymptotic solution, and we can analyze its analytic properties to find excited state TBA equations whose solution we can extend beyond the asymptotic regime.

\subsection{$Y/T/Q$-system and nonlinear integral equations}

The structure of fusion relations between bound state transfer matrix eigenvalues actually relates nicely to a structure that is known as the T system, a system encountered in S. Negro's article \cite{Negro:2016yuu} in a particular model. Let us go over the basic story, avoiding formulas. The T system is a set of equations known as Hirota equations for a set of T functions, functions of the rapidity (momentum) defined on a grid with a border one wider than the Y system on all sides. The identification between the Ys and the Ts admits gauge transformations on the Ts, but in an appropriate gauge the asymptotic Y functions are expressed in terms of asymptotic T functions, for which the (asymptotic) T system becomes precisely equivalent to the fusion relations of the transfer matrix eigenvalues. The T system is a generic rewriting of the Y system however, which applies beyond the asymptotic limit. Its gauge freedom actually proves useful, as one can (try to) shift the analytic properties of the Y functions that we require from the TBA, between the various T functions. Doing so appropriately, we can represent the (typically infinite set of) T functions in terms of a set of much simpler elementary functions known as Q functions with transparent analytic properties. Turning the resulting algebraic equations plus analyticity constraints back into integral equations for these ``fundamental'' variables gives a set of nonlinear integral equations for a finite number of functions, of the general Kl\"umper-Pearce--Destri-de Vega type mentioned above. This hence provides a means of rewriting the TBA equations in a simpler form in these more complicated cases with infinitely many Y functions. In the context of integrability in AdS/CFT these equations are known as the quantum spectral curve \cite{Gromov:2013pga}. S. Negro's article \cite{Negro:2016yuu} discusses that deriving such Y, T, or Q systems and their analytic properties from first principles is possible in particular models. While a highly involved problem, doing so in a particular model would provide a great check on the chain of reasoning involved in the TBA approach (for excited states in models with bound states). For the $\ads$ string first steps in this direction were made in \cite{Benichou:2011ch}.

\section{Conclusion}

The thermodynamic Bethe ansatz is an important technical tool with applications ranging from (but not limited to) describing the thermodynamic properties of one dimensional spin chains to computing the spectra of integrable field theories on a cylinder. In this article we provided an introduction to the basic ideas behind this method, and applied them in a number of illustrative and representative examples. We started from the simplest Bethe ansatz integrable model -- free electrons -- where we introduced the thermodynamic limit and the concept of density of states and holes and their relation via momentum quantization conditions. The stationarity of the free energy in thermodynamic equilibrium resulted in a simple algebraic equation, whose solution gave the famous Fermi-Dirac distribution. We then applied the same ideas with the free particle momentum quantization condition replaced by more complicated Bethe(-Yang) equations, to describe the thermodynamics of the Bose gas, XXX spin chain, and chiral Gross-Neveu model.  These latter two models required us to introduce a string hypothesis describing the possible solutions of the Bethe equations in the thermodynamic limit. The stationarity condition now results in one or or an infinite number of coupled integral equations -- the TBA equations -- for the Bose gas, and XXX spin chain and chiral Gross-Neveu model respectively. We discussed how such infinite sets of TBA equations can be simplified and ultimately reduced to a Y system together with analyticity data, including technical details on integration kernel relations presented in an appendix. We then moved on to using the same ideas to describe the ground state energy of integrable field theories in finite volume via the mirror trick of interchanging space and time, and how these ideas can be adapted and applied to excited states. The Y system structure is the same for all such excited states, and we discussed the analyticity data required to link a Y system to a given model and within that to a given state. We also briefly discussed the basics of and some tips on numerically solving TBA equations. The conceptual background we discussed and applied to our concrete examples make up the essence of the TBA approach, and as such can be applied to (m)any other integrable model(s).

\section*{Acknowledgements}

This article arose out of lectures presented at the YRIS school on integrability organized at Durham university, 6-10 July 2015. I would like to thank the other organizers and lecturers as well as the scientific committee for their efforts, the GATIS network for its support, the participants for their attention and interesting questions, and the Department of Mathematical Sciences at Durham University for its hospitality. I would also like to thank my collaborators on projects related to the TBA for the opportunity to learn about this interesting topic, and many other colleagues for insightful discussions. ST is supported by LT. The work of ST is supported by the Einstein Foundation Berlin in the framework of the research project "Gravitation and High Energy Physics" and acknowledges further support from the People Programme (Marie Curie Actions) of the European Union's Seventh Framework Programme FP7/2007-2013/ under REA Grant Agreement No 317089.

\appendix

\numberwithin{equation}{section}

\section{Integral identities}
\label{app:kernelidentities}

In eqn. \eqref{eq:KQMidentity} of section \ref{sec:TBAtoY}, we claimed that the kernels $K^{MN}$ satisfy
\begin{equation}
\label{eq:KQMidentityapp}
K^{PQ} - (K^{PQ+1}+K^{PQ-1} )\star s = s \, I_{PQ}.
\end{equation}
We also made claims regarding $K^{fQ}=K^{Qf}$, namely
\begin{equation}
\label{eq:KfQidentity}
K^{fQ} - (K^{fQ+1}+K^{fQ-1} )\star s = s \, \delta_{Q1}.
\end{equation}
We can prove these by Fourier transformation. We begin by noting that similarly to the fused XXX momentum of eqn. \eqref{eq:fusedmomentum},
\begin{equation}
S^{fQ}(u) = \frac{u+ iQ}{u- iQ}, \,\,\,\, K^{fQ}(u) \equiv - \frac{1}{2 \pi i} \frac{d}{du} \log S^{1Q}(u) = \frac{1}{\pi} \frac{Q}{Q^2 + u^2}.
\end{equation}
Note that $S^{11} = 1/S^{f2}$, but that we defined the kernels with opposite sign, so $K^{11} = K^{f2}$. Now the Fourier transform of $K^{fQ}$ ($Q\geq1$) is
\begin{equation}
\label{eq:KfQfourier}
\hat{K}^{fQ}(k) \equiv \int_{-\infty}^{\infty} du e^{iku} K^{fQ}(u) = e^{-|k|Q},
\end{equation}
while
\begin{equation}
\hat{s}(k) = \frac{1}{2 \cosh k}.
\end{equation}
In Fourier space, identity \eqref{eq:KfQidentity} is now simply an equality between functions. The identity for $K^{QM}$ similarly follows by its definition as a sum over string states ($K^{f0}=0$)
\begin{align}
K^{QM}(u) & \equiv \sum_{\mbox{strings}} K^{11}= \sum_{\mbox{strings}} K^{f2}\\
& = K^{f(Q+M)}(u) + K^{f(M-Q)}(u) + 2 \sum_{i=1}^{Q-1} K_{M-Q+2i}(u)\\
& = K^{f(Q+M)}(u) + K^{f(|M-Q|)}(u) + 2 \sum_{i=1}^{\mbox{min}(M,Q)-1} K_{|M-Q|+2i}(u)
\end{align}
which we get by combining appropriately shifted numerators and denominators in the product of $S$ matrices underlying these kernels. Its Fourier transform, cf. eqn. \eqref{eq:KfQfourier}, is
\begin{equation}
\hat{K}^{QM} = \sum \hat{K}^{fX} = \coth |k| \left(e^{-|Q-M||k|} - e^{-(Q+M)|k|}\right) - \delta_{Q,M},
\end{equation}
from which eqn. \eqref{eq:KQMidentityapp} follows.

\section{Numerically solving TBA equations}
\label{app:numTBA}

We mentioned in \ref{sec:Bosegas} that we can numerically solve TBA equations by iterations. Let us consider the general form of a TBA equation
\begin{equation}
\log Y_j = \log\left(1+\frac{1}{Y_k}\right) \star K_{kj} + a_j,
\end{equation}
where the $a$ denote a set of driving terms, including for instance the energy term in eqn. \eqref{eq:bosegasTBA}. To solve these equations by iterations, we start with some guess for the Y function(s) as a seed -- the $Y_j^{(0)}$ -- and use these initial functions to compute the right hand side of the TBA equations. We then use this to define the updated $Y_j^{(1)}$, or more generally
\begin{equation}
\label{eq:basiciteration}
\log Y_j^{(n+1)} = \log\left(1+\frac{1}{Y_k^{(n)}}\right) \star K_{kj} + a_j.
\end{equation}
In practice we hope these iterations converge to a stable solution.\footnote{In the case of the free energy for the Bose gas this can be explicitly shown \cite{Yang:1968rm}, but let us simply assume this is ok in general at least as long as we do not choose our initial Y functions too poorly.}
Of course, the trick lies in the technical implementation of this basic concept, which is a bit of an art.

First, a good guess for the initial Y functions will at the very least speed up the process. If we wanted to solve the Bose gas equations \eqref{eq:bosegasTBA}, for instance, in a low temperature regime a good guess would be $\epsilon^{(0)}(p) = E(p)$. Second, depending on the details of the equations and kernels, nothing guarantees that eqn. \eqref{eq:basiciteration} will converge fastest. For instance, it may be advantageous to consider
\begin{equation}
\label{eq:variediteration}
\log Y_j^{(n+1)} = x\left(\log\left(1+\frac{1}{Y_k^{(n)}}\right) \star K_{kj}  + a_j\right) + (1-x)\log Y_j^{(n)},
\end{equation}
for some $0<x\leq1$, cf. e.g. section 2 of \cite{Dorey:1996re}. This is mostly useful if we need to run similar equations many times, since finding a suitable value for $x$ through experimentation takes time as well. Third, the convolution computations can typically be sped up by means of (fast) Fourier transform (FT), i.e. we compute the convolution $f \star g$ as $\mbox{FT}^{-1}(\mbox{FT}(f)\mbox{FT}(g))$.\footnote{There are nice exercises with solutions illustrating this as part of the 2012 edition of the Mathematica summer school on theoretical physics available on the web \cite{janikmathematicaschool}.} Alternatively we could try to solve the equations in Fourier space directly, for example by using a multidimensional version of Newton's method at a discrete set of values in the Fourier variable. It may in fact be useful to use Newton's method when iterating in whatever form, see e.g. \cite{Takahashi:2001uq}: rather than updating as $Y_M^{(n+1)} = Y_M^{(n)} + \Delta_M^{(n)}$, where $\Delta_M^{(n)}$ denotes the error of the solution at iteration $n$, we could update in the direction of greatest linear improvement, i.e. as  $Y_M^{(n+1)} = Y_M^{(n)} + \xi_M^{(n)}$ where $\xi_M^{(n)}$ solves $(\delta_{M}^N - \partial \mbox{RHS}_M(Y^{(n)})/\partial Y_N )\xi^{(n)}_N = \Delta_M^{(n)}$ and $\mbox{RHS}_M(Y^{(n)})$ denotes the right hand side of the TBA equations at iteration $n$.

Regarding the technical implementation of these convolutions and sums, on a computer we cannot work with infinitely many Y functions or integrals over the whole real line. This means that in case of infinitely many Y functions we will have to cut them off at some point, and in any case the integrals will need to be done through some  discretized finite interval. Regarding this first point, typically the Y functions for bound states fluctuate less and give smaller contributions to the free energy as the bound state size grows. Consider for instance the constant asymptotics of $Y_Q \sim Q(Q+2)$ that we mentioned in section \ref{sec:YsystoTBA}, meaning that $\log(1+1/Y_Q)$ decreases with $Q$, unless its relative fluctuations grow in $Q$, which would be odd. So for practical numerical purposes it may suffice to keep only e.g. the first ten Y functions, unless self-consistency checks based on these first ten indicate that the contributions of higher Y functions are not negligible with regard to the desired accuracy. Importantly, we should not simply drop the other Y functions altogether, but rather add for instance the contribution of their constant asymptotics. This brings us to the second point, integrating over a finite interval. Since we need to cut off the integration domain in some fashion, we need to take care of the asymptotics anyway: cutting the integration domain off at a fixed value means we will introduce a boundary error of order of the asymptotic value at the extrema of the external parameter in the convolution.\footnote{$K(u-v)$ with $u\sim v$ is of order $K(0)$ which is a relevant scale.} To reduce this error to an acceptable value we can subtract the equation satisfied by the constant asymptotics, i.e. we solve
\begin{equation}
\log Y_j = \log A_j + \log\left(1+\frac{1}{Y_k}\right)/\left(1 + \frac{1}{A_k}\right) \star K_{kj} + a_j,
\end{equation}
where $A_i$ denotes the asymptote of $Y_i$, which here we assumed to solve the TBA equations with $a_i=0$. If there are constant nonzero asymptotics in the game, subtracting them is also essential if we wish to Fourier transform. Nonzero constants Fourier transform to delta functions which cannot be reliably implemented numerically. Put differently, functions with constant nonzero asymptotics are not square integrable on the line, so cannot be Fourier transformed in the traditional sense. If we subtract the asymptotics, however, we can readily Fourier transform the fluctuations of interest.

The discussion in this appendix applies equally well to simplified TBA equations -- nothing referred to the canonical form of eqn. \eqref{eq:basiciteration} -- which importantly are typically faster for numerical purposes as they do not involve infinite (large) sums, but nearest neighbour couplings instead. As discussed we need to be careful about the asymptotics we subtract: in contrast to the canonical equations there are many constant solutions of the basic simplified TBA equations, and we have to choose the one appropriate for our physical situation.

\bibliographystyle{jhep}

\bibliography{stijnsbibfile}


\end{document}